\documentclass[11pt,preprint]{aastex}
\usepackage{amsmath}
\usepackage{amssymb}
\usepackage{epsfig}

\newcommand{\cms}{\,{\rm cm$^{-2}$}\,}
\newcommand{\cmc}{\,{\rm cm$^{-3}$}\,}
\newcommand{\kms}{\,{\rm km\,s$^{-1}$}\,}

\newcommand{\kel}{\,{\rm K\ }}
\newcommand{\etal}{{ et~al.~}}
\newcommand{\expunit}{\,{\rm photons\,s$^{-1}$\,cm$^{-2}$\,arcsec$^{-2}$}\,}

\newcommand{\sbunit}{\,{\rm counts\,cm$^{-2}$\,bin$^{-1}$}\,}
\newcommand{\ergs}{\,{\rm erg\,s$^{-1}$}\,}
\newcommand{\ergscm}{\,{\rm erg\,s$^{-1}$\,cm$^{-2}$}\,}
\newcommand{\Ms}{\,M_\odot\,}
\newcommand{\Zs}{\,Z_\odot\,}

\shorttitle{{\it Chandra} Observations of NGC~4552}
\begin{document}


\title{{\it CHANDRA} OBSERVATIONS OF GAS STRIPPING 
 IN THE ELLIPTICAL GALAXY NGC~4552 IN THE VIRGO CLUSTER}

\author{M. MACHACEK, C. JONES, W. R. FORMAN, AND
P. NULSEN$^\dagger$\altaffiltext{$\dagger$}{on leave from the University of Wollongong}}
\affil{Harvard-Smithsonian Center for Astrophysics \\ 
       60 Garden Street, Cambridge, MA 02138 USA
\email{mmachacek@cfa.harvard.edu}}

\begin{abstract}

We use a $54.4$\,ks {\it Chandra} observation  to 
study ram-pressure stripping  
in NGC~4552 (M89), an elliptical galaxy in the Virgo Cluster.
{\it Chandra} images in the 
$0.5-2$\,keV band show a sharp leading edge in the surface brightness
 $3.1$\,kpc north of the galaxy center, a cool 
($kT =0.51^{+0.09}_{-0.06}$\,keV) 
tail with mean density $n_e \sim 5.4 \pm 1.7 \times 10^{-3}$\cmc  
extending $\sim 10$\,kpc to the south of the galaxy, and  two
$3-4$\,kpc horns of emission extending southward away from the 
leading edge. These are all features characteristic of supersonic 
ram-pressure stripping of galaxy gas, due to NGC~4552's motion 
through the surrounding Virgo ICM.  Fitting the surface brightness profile 
and spectra across the leading edge, we find the galaxy gas 
inside the edge is cooler ($kT = 0.43^{+0.03}_{-0.02}$\,keV) 
and denser  ($n_e \sim 0.010$\cmc) 
than the surrounding Virgo ICM ($kT = 2.2^{+0.7}_{-0.4}$\,keV and 
$n_e = 3.0\pm 0.3 \times 10^{-4}$\cmc).
The resulting pressure ratio between the free-streaming ICM and
cluster gas at the stagnation point is
$\sim 7.6^{+3.4}_{-2.0}$ for galaxy gas metallicities of 
$0.5^{+0.5}_{-0.3}\,\Zs$, which suggests that NGC~4552 is 
moving supersonically through the cluster  with a velocity  
$v \sim 1680^{+390}_{-220}$\kms (Mach $2.2^{+0.5}_{-0.3}$) at  
an angle $\xi \sim 35^{\circ} \pm {7^\circ}$ 
towards us with respect to the plane of the sky.
\end{abstract}



\section{INTRODUCTION}
\label{sec:introduction}

Hierarchical models of structure formation, in which galaxies formed 
at high redshift and rapidly coalesced into groups and clusters 
through accretion and mergers, have become a compelling paradigm for 
understanding galaxy evolution. In such dynamically rich environments,
 galaxies are subject to an array of physical processes 
that affect their evolution. 
These physical processes form two broad classes: 
(1) tidal interactions such as 
those induced by major mergers (e.g., Lavery \& Henry 1988), off-axis galaxy 
collisions (M{\"u}ller \etal 1989), galaxy harassment 
(Moore \etal 1996) or galaxy fly-bys near the core of the 
cluster potential (Byrd \& Valtonen 1990), and (2) gas-gas 
interactions, notably ram pressure by the intracluster gas 
(ICM) on the galaxy's interstellar medium 
(ISM), due to the galaxy's motion through the surrounding ICM 
(Gunn \& Gott 1972), or ISM-ISM interactions produced in 
galaxy-galaxy collisions (Kenney \etal 1995). These gas-dynamical 
processes may be enhanced by turbulence and viscous effects 
(Nulsen 1982; Quilis \etal 2000) or inhomogeneities and bulk motions 
in the ICM gas (Kenney \etal 2004). 
Combes (2004) and references therein provide a recent 
overview. No model for galaxy evolution can be complete without a 
detailed understanding of the gas-dynamics of these interactions. 
Similarly no model for the evolution of the intracluster medium  can be
complete without understanding the feedback of gas and energy from the
galaxy into the ICM by these gravitational and hydrodynamical 
stripping processes. 

The actions of tidal forces are identified by the appearance of 
disturbed stellar morphologies, such as stretched stellar tails 
(see e.g. Gnedin 2003, Vollmer 2003); while the characteristic
signatures of hydrodynamic stripping (see e.g. Stevens \etal 1999; 
Toniazzo \& Schindler 2001; Acreman \etal 2003 and
references therein) 
are imprinted
on the structure of the hot X-ray emitting gas in and near the galaxy.
Notably the ram-pressure experienced by galaxies moving through 
the dense ICM can produce sharp surface brightness edges 
(``cold fronts'') along the leading interface between the galaxy 
and the ICM (Acreman \etal 2003; Machacek \etal 2005a). These 
cold fronts are contact discontinuities where a sharp rise in 
surface brightness (gas density) is accompanied by a corresponding drop 
in gas temperature, and are the galaxy-sized analogues of similar       
subcluster-scale features identified near the cores of rich clusters
(see e.g. Markevitch \etal 2000, Vikhlinin \etal 2001; 
Mazzotta \etal 2001). 
 The higher pressure found in the cold front 
compared to that in the cluster gas is balanced by the ram pressure 
caused by the motion of the cloud (in our case, galaxy) 
through the surrounding ICM. Tails of stripped galaxy gas 
 have been found to extend $10$ to $200$\,kpc behind the galaxy 
before fading into the background emission of the ambient cluster
gas (see e.g. Forman \etal 1979, White \etal 1991, Rangarajan \etal 1995 
for M86; Irwin \& Sarazin 1996, Biller \etal 2004 for NGC~4472; 
Sakelliou \etal 1996 for 4C34.16; Wang \etal 2004 for C153; 
Machacek \etal 2005a, Scharf \etal 2005 for 
NGC~1404). Detailed studies of these X-ray features in nearby
systems (see, for example, Biller \etal 2004; 
Machacek \etal 2005a, 2005b; Sun \& Vikhlinin 2005; 
Sun, Jerius \& Jones 2005), made possible by the high angular 
resolution of {\it Chandra} and {\it XMM-Newton}, not only reveal 
the nature of the gas-dynamics of the stripping process, but also are 
one of the few ways to constrain the three dimensional motion of the 
galaxy as it passes through the group or cluster core  
(Merrifield 1998; Dosaj \etal 2002). 

NGC~4552 
($\alpha = 12^h35^m39.8^s$,$\delta =12^\circ33'23''$, J2000) is 
an elliptical galaxy located $72'$ east of M87 
in subcluster A of the Virgo cluster. 
Measurement of NGC~4552's distance modulus using surface brightness
fluctuations (Tonry \etal 2001) places it at nearly the same 
luminosity distance as M87, while NGC~4552's low 
line of sight velocity ($v_r = 340 \pm 4$\kms, NED; Smith \etal 2000) 
compared to that of M87 
($v_r = 1307 \pm 7$\kms, NED; Smith \etal 2000) 
implies that NGC~4552 is moving supersonically through the Virgo ICM  
with at least $\Delta v_r = -967 \pm 11$\kms towards us relative to M87. 
 Thus NGC~4552 is a likely candidate for ram-pressure stripping by 
the cluster gas. 

Since NGC~4552 is a member of the large sample of nearby elliptical 
galaxies used to establish correlations between X-ray emission and 
other ISM tracers (see e.g. O'Sullivan \etal 2001 and references
therein), its global X-ray properties are well studied.
Early observations with the {\it Einstein} Observatory 
(Forman, Jones \& Tucker 1985; Canizares, Fabbiano \& 
Trinchieri 1987; Roberts \etal 1991; 
Kim, Fabbiano, \& Trinchieri 1992; Eskridge, Fabbiano, \& Kim 1995) 
focused on the measurement of NGC~4552's total X-ray luminosity 
to establish the $L_{\rm X}$ --$L_{\rm B}$ relation. Subsequent 
{\it ROSAT} observations of NGC~4552 were used to study the 
possible dependence of this relation on environment 
(Brown \& Bregman 2000; O'Sullivan \etal 2001). Fits to the mean 
X-ray spectrum for NGC~4552 using {\it ROSAT} 
resulted in  mean temperatures for the X-ray gas in 
the galaxy of $\sim 0.5 - 0.8$\,keV  depending on the spectral model
(Davis \& White 1996; Brown \& Bregman 2000; Matsushita 2001). 
Using {\it ASCA} data, Matsushita, Ohashi, \& Makishima (2000)
modelled the mean spectrum of NGC~4552 using variable abundance 
thermal plasma models 
including an additional bremsstrahlung component in the spectral models to 
account for the contribution of unresolved 
 point sources in low spatial resolution data. They found mean temperatures 
$\sim 0.6$\,keV for NGC~4552, in agreement with the {\it ROSAT} 
studies, but with generally higher abundances ($\gtrsim 0.4\,\Zs$).
Most recent studies, across a wide range of wavelengths, 
have concentrated on the nuclear properties of NGC~4552. A brief
 review of prior work on the nucleus can be found in 
Machacek \etal (2006; hereafter known as Paper II), where we
 present our results from Chandra observations of the central regions 
of NGC~4552 that show shocks close to the center of NGC~4552 
that were produced by recent nuclear activity.

In this paper we use {\it Chandra} X-ray 
observations to focus on the properties of gas in the outer 
regions of NGC~4552 where hydrodynamic gas stripping is occuring.
Our discussion is organized as follows:  
In  \S\ref{sec:obs} we describe the observations and our 
data reduction and processing procedures.
In  \S\ref{sec:ramresult} we  present the background 
subtracted, exposure corrected image of NGC~4552 showing the 
prominent X-ray emission features of the system: 
a sharp surface brightness edge to the north, 
'horns' of emission extending away to the south 
from the surface brightness edge, a ram pressure stripped X-ray tail,  
a bright nucleus, and bright ringlike features  (discussed in 
detail in Paper II) near the galaxy center. 
We  then discuss our analysis method  and main results, including the 
determination of the gas density and temperature, for gas 
stripping from the outer regions of the galaxy.  
We use these results 
to constrain the velocity of NGC~4552 through the Virgo ICM. 
In \S\ref{sec:conclude} we summarize our results. 
Unless otherwise indicated all errors correspond to  $90\%$ 
confidence levels and coordinates are J2000. Taking the distance to 
the dominant elliptical 
M87 as representative of the distance to subcluster A of the Virgo
Cluster containing NGC~4552, the luminosity distance to the cluster
is $16.1 \pm 1.1$\,Mpc (Tonry \etal 2001)  
and $1''$ corresponds to a distance scale of $77$\,pc.


\section{OBSERVATIONS AND DATA REDUCTION}
\label{sec:obs}

For our analysis, we used a $54.4$\,ks observation of the elliptical galaxy
NGC~4552 in Virgo taken with {\it Chandra} 
 on 2001 April 22 using the Advanced CCD Imaging Spectrometer array 
(ACIS, Garmire \etal 1992; Bautz \etal 1998) with ACIS-S (chip S3) at
the aimpoint. The focal plane temperature of the instrument was
$-120^\circ$\,C throughout the observation.
The data were analyzed using the standard X-ray processing packages, 
CIAO version $3.1$, FTOOLS and XSPEC version $11.2$. 
Filtering removed events with 
bad grades ($1$, $5$, and $7$) and those with significant flux in 
border pixels of the  $5 \times 5$ event islands (VFAINT mode), as well
as any events falling on hot pixels. Use of VFAINT mode improves
background rejection by as much as a factor of $3$ in the S3 chip
for soft X-ray energies ($\sim 0.3$\,keV) important to this
observation. The data were reprocessed and response files created   
using the most recent gain tables and instrumental corrections. 
These included correcting for the time-dependent declining efficiency of the 
ACIS detector due to the buildup of contaminants on the optical filter
(Plucinsky \etal 2003), which is important at energies below $1.5$\,keV,
and for the slow secular drift (tgain) of the average PHA values for photons
of fixed energy.\footnote{see Vikhlinin \etal in
http://cxc.harvard.edu/contrib/alexey/tgain/tgain.html}
Periods of anomalously high background (flares) were identified and
 removed from the data, along with periods of anomalously low 
count rates at the beginning and end of the
observation. This resulted in a useful exposure time of $51,856$\,s.

Backgrounds for use in the imaging analyses and spectral measurements 
 of the Virgo Cluster
gas were created from the $450$\,ks period D source free 
dataset (aciss\_D\_7\_bg\_evt\_271103) appropriate for the 
date of observation and instrument configuration.\footnote{see 
http://cxc.harvard.edu/contrib/maxim/acisbg }. 
Identical cleaning, energy and spatial
filters were applied to source and background data throughout.
We checked the standard normalization of the source free background, 
set by the ratio of the exposure times, by comparing count 
rates between the source and background files in the $9.5 - 12$\,keV 
energy range, where particle background dominates. We  found the 
standard normalization was $5\%$ too high and renormalized the source
free data by a  factor of $0.95$ to correct for this difference.
Point sources were identified in the $8'.4 \times 8'.4$ field of the 
S3 chip in the $0.3-10$\,keV energy band using a multiscale 
wavelet decomposition algorithm set with a $5\sigma$ detection
threshold. The resulting $132$ source identifications
were excluded from the spectral and surface brightness
analyses that follow.

\section{RESULTS: RAM PRESSURE STRIPPING OF NGC~4552}
\label{sec:ramresult}

\begin{figure}[t]
\begin{center}
\includegraphics[height=2.7285in,width=3in]{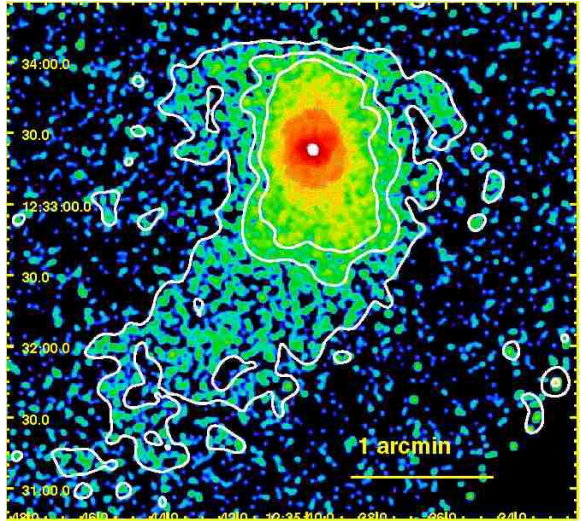}
\caption{\footnotesize{{\it Chandra} image of the $0.5-2$\,keV diffuse emission in 
NGC~4552 in the Virgo Cluster. Point sources have been removed
and the image has been background subtracted, exposure corrected and 
smoothed with a Gaussian kernel with $\sigma = 1''$. 
Contours denote X-ray surface brightness levels of 
$0.4,\,1,\,2 \times 10^{-8}$\expunit, respectively.  
The X-ray surface brightness exhibits a sharp 
leading edge $40''$($3.1$\,kpc) to the north of
the galaxy center, horns of emission extending southeast and
southwest of the leading edge, and a $\sim 2'$ ($\sim 10$\,kpc) tail of
emission to the south-southeast. The bright inner rings 
of emission  $\sim 17''$($1.3$\,kpc) from the galaxy's nucleus are 
discussed in Paper II. 
North is up and east is to the left. }}
\label{fig:image}
\end{center}
\end{figure}

\begin{figure}[t]
\begin{center}
\includegraphics[height=3in,width=3in,angle=0]{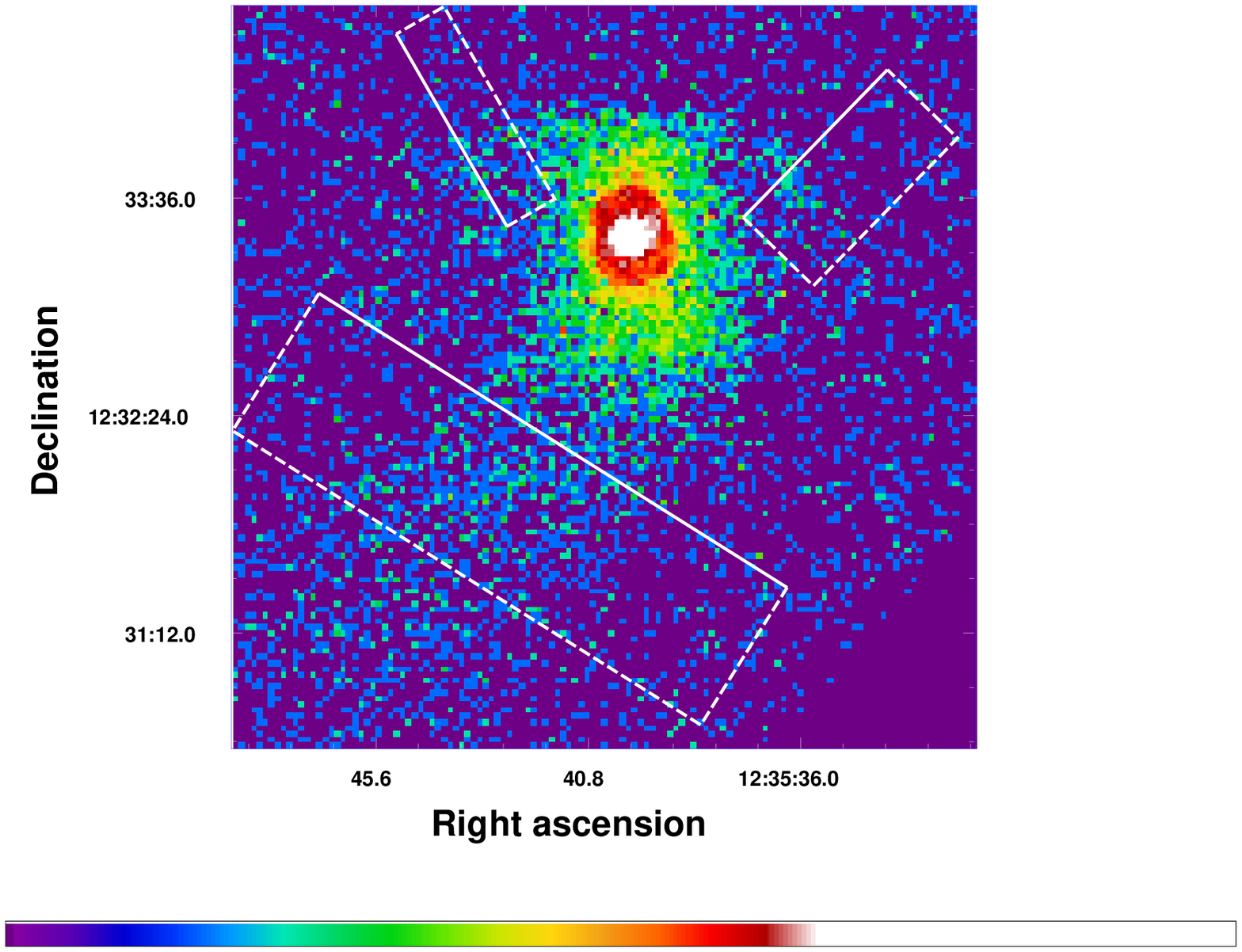}
\hspace{0.3cm}
\includegraphics[height=3in,width=3in,angle=0]{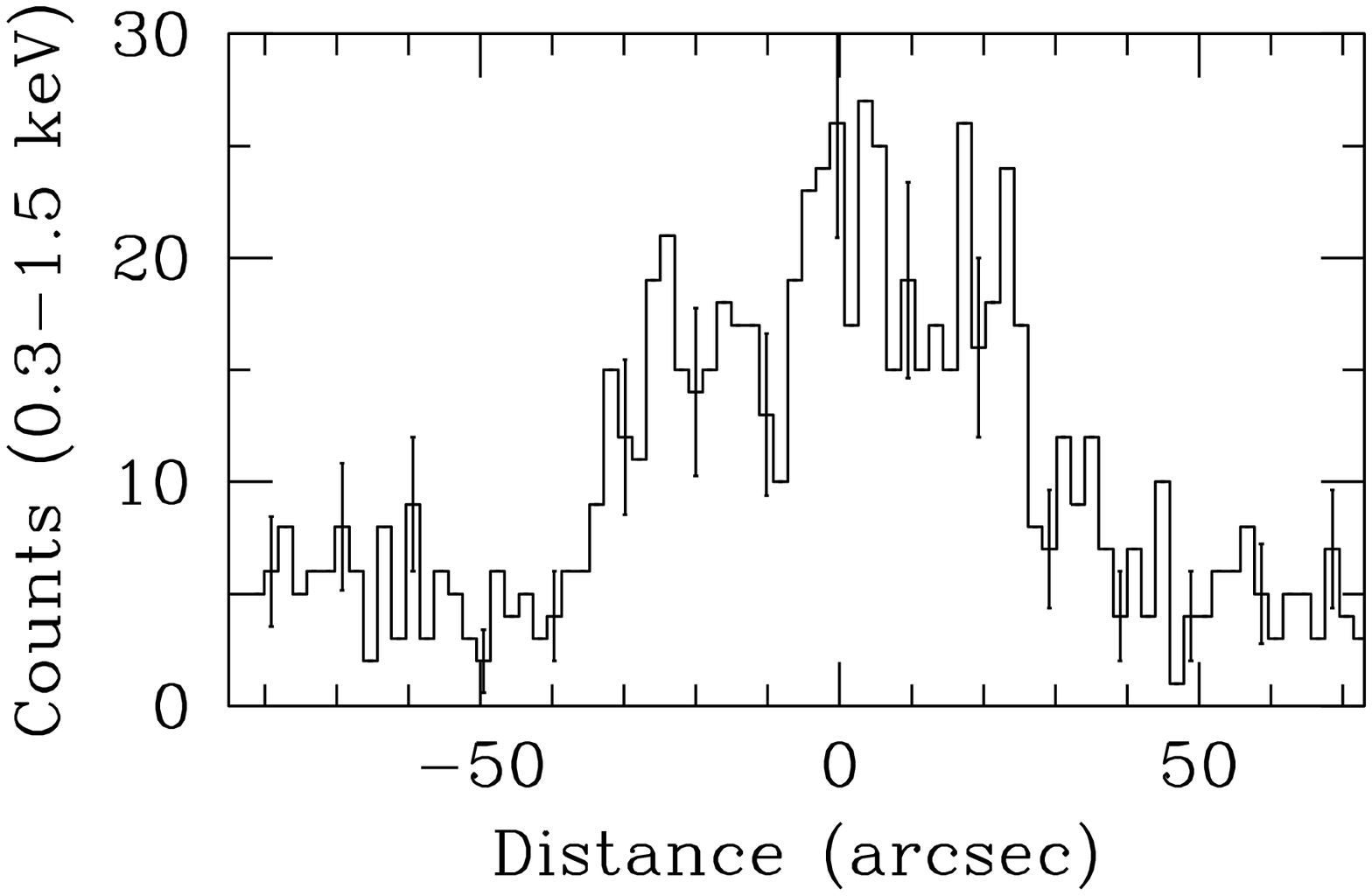}
\vspace{0.3cm}
\includegraphics[height=3in,width=3in,angle=0]{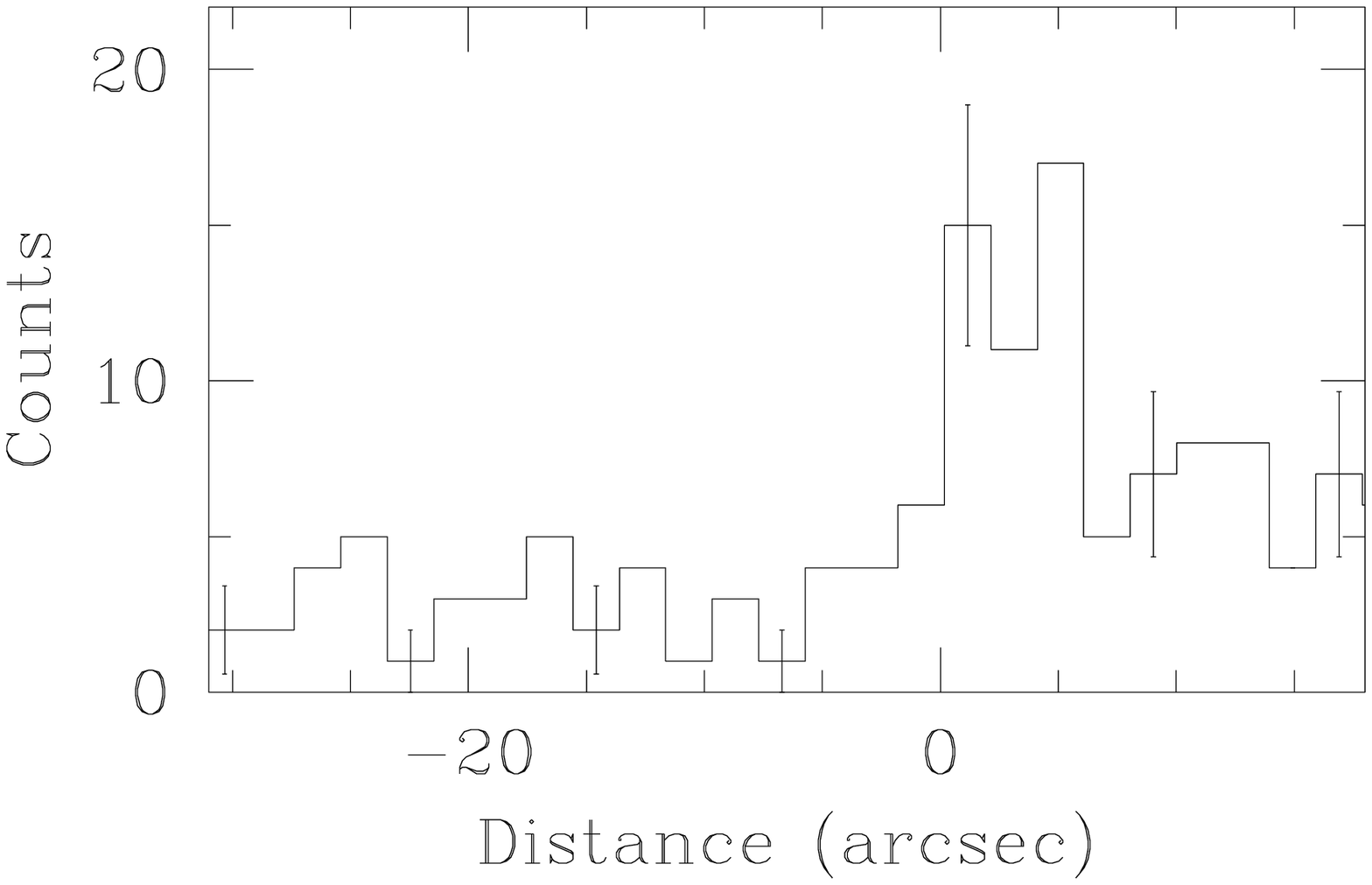}
\hspace{0.3cm}
\includegraphics[height=2in,width=2in,angle=0]{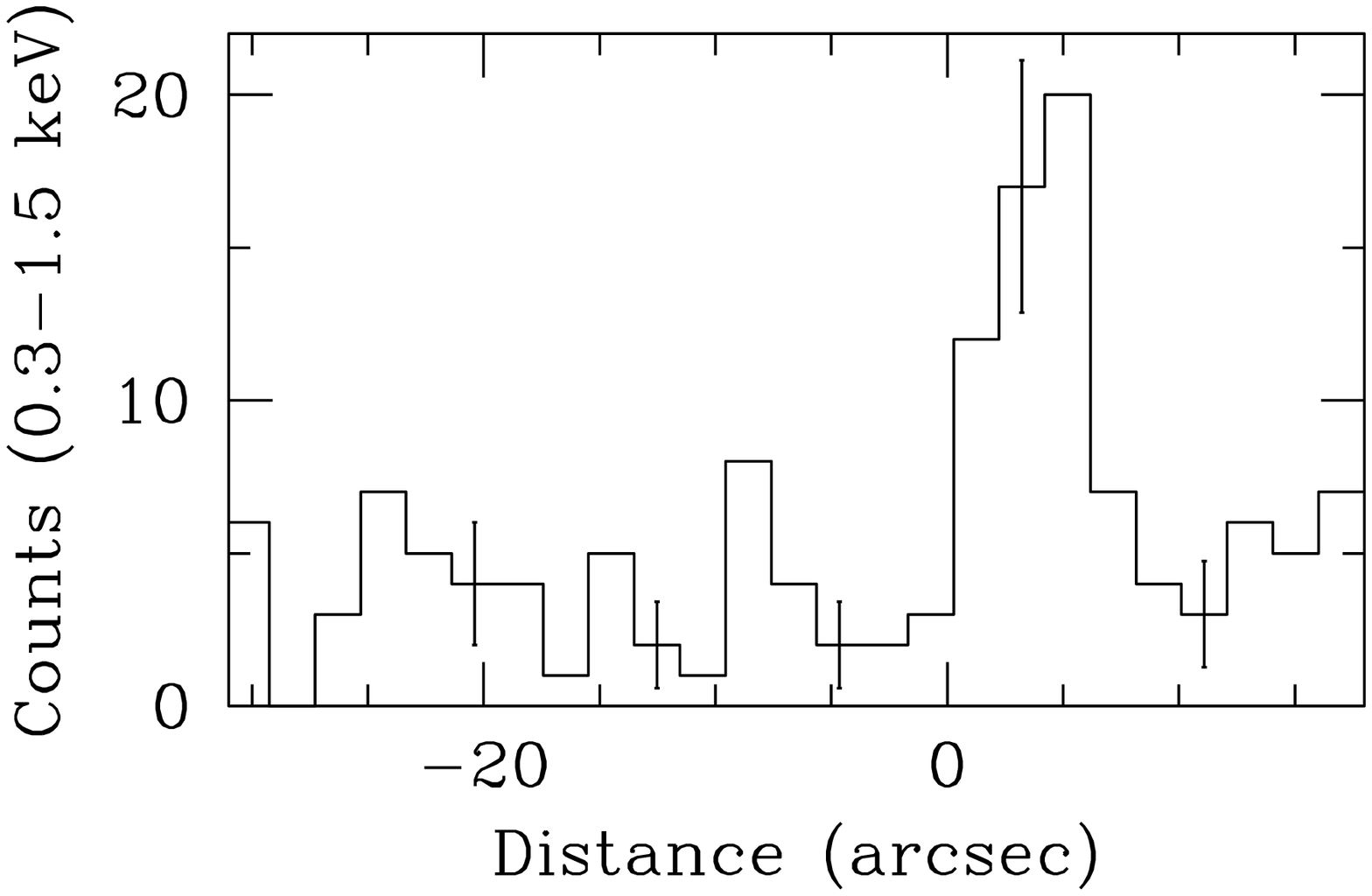}
\caption{\footnotesize{ ({\it top left}) $0.3-1.5$\,keV Chandra image of NGC~4552
superposed with projection regions across the eastern and western
horns and the tail. The image has $2'' \times 2''$ pixels. Point
sources have been removed using CIAO tool {\it dmfilth}. The resulting 
projections across the tail, eastern horn and western horn are shown
in the top right, lower left and lower right panels, respectively. 
Taking the background from the cluster (left) side of the horn
features gives probabilities of $\lesssim 10^{-14}$ 
that the excess counts could be due to 
a fluctuation of the background, while using 
the higher values from the galaxy (right) side of the projection 
gives probabilities of $\lesssim 10^{-5}$. }}
\label{fig:hehwtsig}
\end{center}
\end{figure}

In Figure \ref{fig:image} we present the $0.5-2$\,keV {\it Chandra}
image of the diffuse emission in NGC~4552 overlaid with X-ray surface 
brightness contours showing the X-ray features 
of interest for our analysis. Point sources were removed from the 
image and the point source regions filled with the local 
average emission level using CIAO tool {\it dmfilth}. The image was 
then background subtracted, corrected for
telescope vignetting and detector response using 
exposure maps created with standard CIAO tools. 
First, two narrow band ($0.5-1$ and $1-2$\,keV) images were background
subtracted and exposure corrected using monoenergetic instrument maps
of $0.9$ and $1.5$\,keV, respectively. The fluxed narrow band images 
were then summed and smoothed with a $1''$ Gaussian kernel to produce 
the image in Figure 1. 
We chose  conservative Gaussian smoothing for our images to minimize
potential numerical artifacts from the smoothing algorithm. 
 The $1''$ smoothing scale represents a 
compromise between the need for high spatial resolution to map sharp 
or narrow features and the need to smooth over larger scales to 
highlight faint extended emission features.
We see a sharp, flattened discontinuity (edge) in the surface brightness 
$3.1$\,kpc ($40''$) north of NGC~4552's center. 
Horns of emission extend $\sim 3-4$\,kpc ($\sim 40''-50''$) to the 
southeast and southwest away from each side of the flattened  edge. 
An X-ray tail extends   
$\sim 10$\,kpc ($\sim 2'$) to the south-southeast before becoming 
indistinguishable from the ambient ICM. In Fig. \ref{fig:hehwtsig} 
we show projections across the horns and tail in a $0.3-1.5$\,keV image
of the diffuse emission (binned to pixel size $2'' \times 2''$ but
without background subtraction), 
demonstrating that these features are highly significant.
\clearpage
\clearpage
\begin{figure}[t]
\begin{center}
\includegraphics[height=2.65in,width=3in,angle=0]{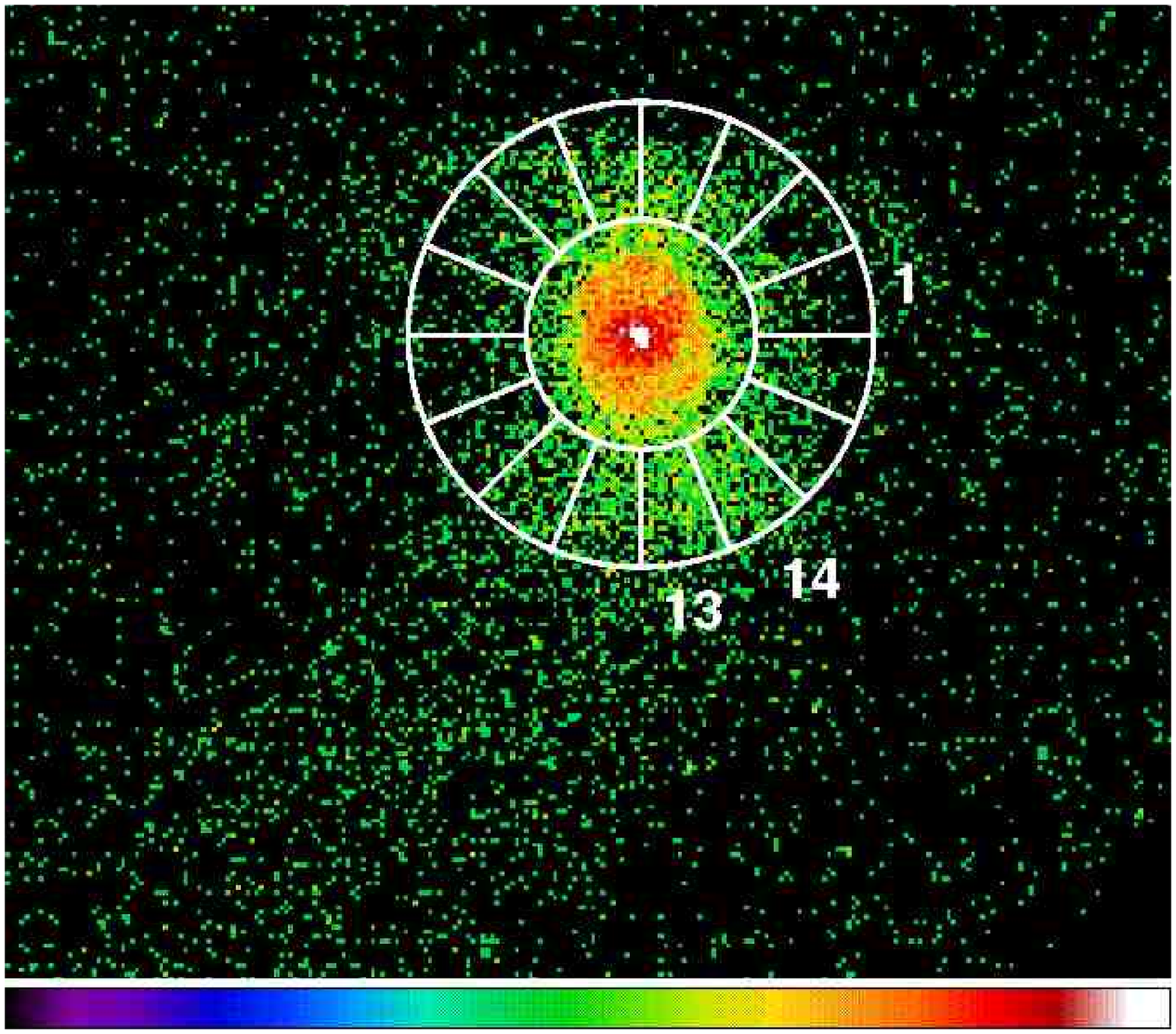}
\includegraphics[height=3in,width=3in,angle=0]{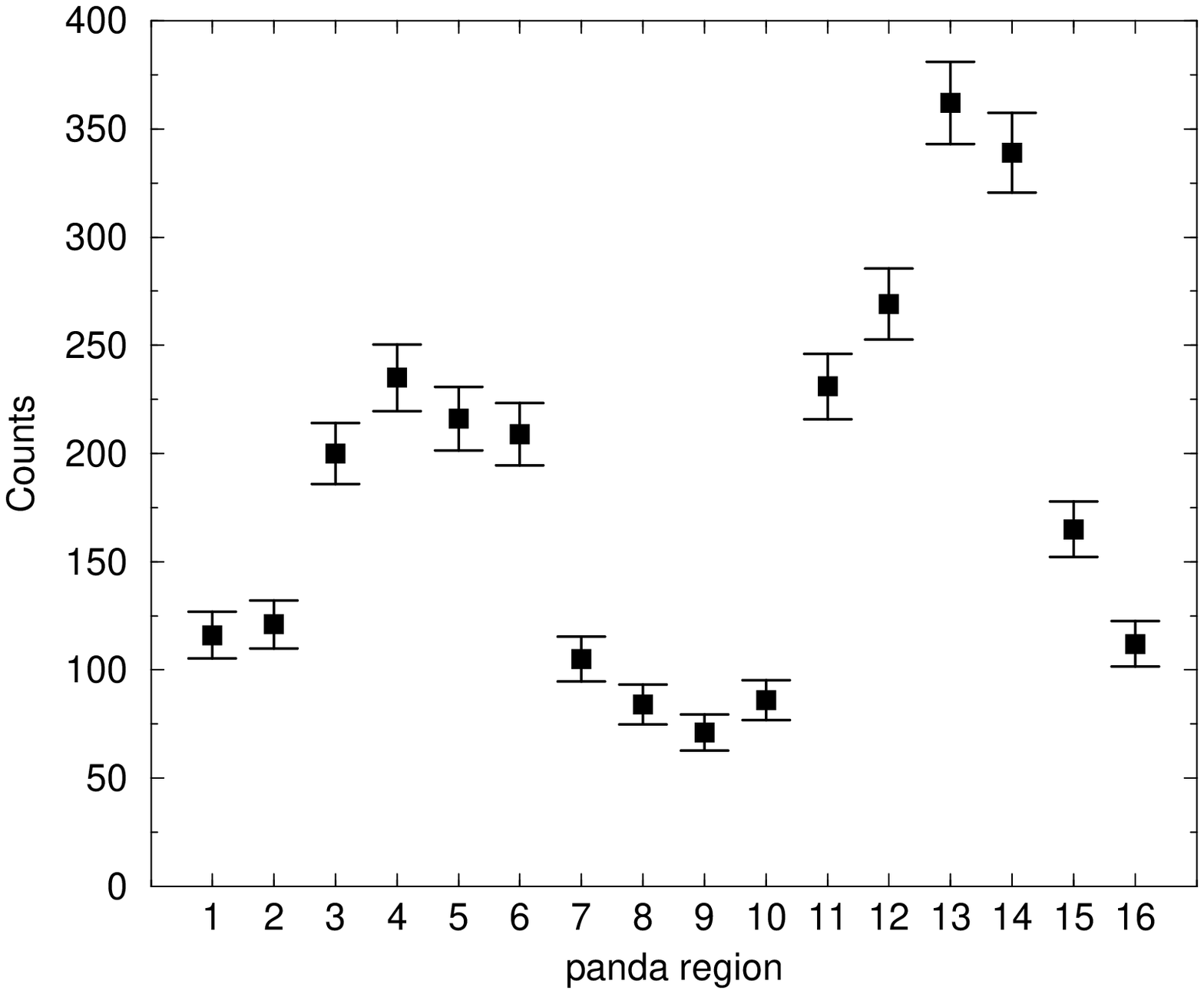}
\caption{\footnotesize{ ({\it left}) $0.3-1.5$\,keV Chandra image of NGC~4552
with $16$ equal angular sectors superposed. Sectors are numbered
counter-clockwise from $1$. The image is binned in  $1'' \times 1''$
pixels. Point sources have been removed using CIAO tool {\it dmfilth},
but no background subtraction has been performed. ({\it right})
Counts in each sector obtained from the image in the left panel. Note that
sectors $3-6$ correspond to the regions containing the leading edge
and sectors $11-14$ are regions in the galaxy near the beginning of
the tail. The peak emission occurs in sectors $13-14$ at the location
of the southern lump (SL).
}}
\label{fig:panda}
\end{center}
\end{figure}

We see in Figure \ref{fig:image} an enhanced region of X-ray emission 
in the southwest quadrant of the galaxy, near the beginning of the
tail, that we denote as the southern lump SL. In Figure
\ref{fig:panda} we  
show the $0.3-1.5$\,keV emission in the galaxy 
for radii $24'' \leq r \leq 48''$   
in $16$ equal sectors numbered counterclockwise from $1$. As shown in
the right panel of Figure \ref{fig:panda}, X-ray emission
from the outer radii of the galaxy is very asymmetric.
Enhanced soft emission occurs both in the northen galaxy regions 
(sectors $3-6$), behind the leading edge of the galaxy, and in the 
southern galaxy regions (sectors $11-14$) near the beginning of the tail. 
The peak in the soft emission for these regions occurs in the
southwest quadrant of the galaxy in sectors $13-14$, at the position
of the southern lump (SL). Thus we see that the southern lump SL, as
well as the horns and the tail, is also highly significant.  

Disturbed X-ray morphologies may arise from tidal interactions 
caused by a recent or ongoing merger, winds of outflowing material
due to a recent starburst, or outflows from a central AGN, 
as well as from ram pressure stripping 
of the galaxy ISM due to the motion of the galaxy through the ambient 
Virgo ICM. For NGC~4552 these first three  
scenarios are disfavored. Tidal (gravitational) interactions act on 
both stars and gas, producing characteristic stellar tidal streams or 
tails (Gnedin 2003; Combes 2004). There is no evidence for disturbed 
morphology in the stellar distribution of NGC~4552 or for significant 
extragalactic stellar light that might result, if the observed
features were of tidal origin (Bettoni \etal 2003). 

A second possibility, given the classification of NGC~4552
as a transition nucleus (Ho \etal 1997), is that the distorted 
X-ray morphology may be due to winds from a compact nuclear starburst.
However, the transition nucleus classification for NGC~4552 is weak,
due to $30-50\%$ uncertainties in the measurement of the H$\beta$ 
line emission (Ho \etal 1997), and the radio spectrum 
of NGC~4552 is inconsistent with that expected
from such a  starburst. The nucleus of NGC~4552  exhibits a 
compact core radio source with a flat spectrum, whose peak brightness 
temperature of $7.8 \times 10^8$\kel (Filho \etal 2004) is several 
orders of magnitude above the $10^5$\kel  upper limit 
(Condon \etal 1991) for the brightness temperature of a compact
nuclear starburst. In addition, starburst winds tend to produce 
conical outflows with associated optical line (H$\alpha$) emission 
from the interaction
between the supernova driven ejecta and the surrounding ISM 
(see, e.g., Strickland \etal 2000 for NGC 253 and Cecil,
Bland-Hawthorn \& Veilleux 2002 for NGC 3079), which are not seen in 
NGC~4552. 

A third possibility is that AGN outbursts could be responsible
for the distorted X-ray morphology seen at large radii 
in NGC~4552. 
The nucleus of NGC~4552 does harbor a supermassive black hole (Filho 
\etal 2004) and does undergo outbursts, as evidenced by the 
ring structures seen in Figure \ref{fig:image} 
and analyzed in detail in Paper II.
However, the X-ray signatures of such
outbursts are residual bright-rimmed cavities and buoyant bubbles,  
as found in  M87 (e.g., Forman \etal 2005), and/or X-ray edges 
corresponding to shocks driven into the ambient medium by the 
outburst (e.g., Fabian \etal 2003 for NGC~1275 in Perseus and  
Nulsen \etal 2005 for Hydra A). This is quite different from the 
`flattened leading edge - trailing tail' morphology seen in 
Figure \ref{fig:image}. 

In contrast, the qualitative correspondence between the main 
X-ray features shown in Figure \ref{fig:image}, i.e. 
the leading edge, horns, and tail, and those 
found in simulations (e.g. see Stevens \etal 1999; 
Toniazzo \& Schindler 2001; Acreman \etal 2003) 
of ram-pressure stripping of elliptical galaxies 
moving through surrounding cluster gas is striking.
However, since none of these simulations
model the specific galaxy, orbital, or cluster characteristics of 
NGC~4552 in Virgo, only a qualitative, not quantitative, comparison 
of these simulations to our observation should be made. 
For supersonic stripping, the galaxy ISM is initially pushed back and 
fanned out (Stevens \etal 1999, Model 1b, Figure 2; 
Toniazzo \& Schindler 2001, Figure 3; Acreman \etal 2003, Figure 2c,
$t=1.6$\,Gyr slice) causing the edge between the galaxy gas and the 
ICM to flatten, as is seen for NGC~4552 (Fig. \ref{fig:image}). 
This sharp surface brightness discontinuity,
$3.1$\,kpc north of the galaxy's center, coupled with a gas tail in
the opposite direction, fixes the galaxy's  direction of motion in 
the plane of the sky. Irregular filaments of stripped material 
also are seen in simulation images extending back from the 
leading edge, similar to the 'horns' we observe in NGC~4552. These filaments 
signal the onset of Kelvin-Helmholtz instabilities, where stripping is
occurring at the boundary between galaxy and ICM gas.
In the  simulation images, galaxy gas, once stripped and decelerated,
forms a distinctive tail of emission behind the galaxy, again similar to 
what we see in Figure \ref{fig:image}. The three-dimensional simulations
of Toniazzo \& Schindler (2001) demonstrate that the tail is in
general not axisymmetric with respect to the direction of motion, but may
appear angular or curved, as is also found in our image.
Thus the most likely explanation for the origin of the features shown  
in Figure \ref{fig:image} is ram-pressure stripping of galaxy gas 
due to the motion of NGC~4552 through the Virgo Cluster ICM.
In order to investigate quantitatively these complex  gas-dynamical 
processes and test our ram-pressure-stripping hypothesis, we first 
determine the spectral properties and densities of hot gas in the 
surrounding ICM and in NGC~4552's main emission features.

\subsection{ Gas Temperatures and Abundances in the Outer Regions}
\label{sec:ramspectra}

\begin{figure}[t]
\begin{center}
\includegraphics[height=2.95in,width=4in]{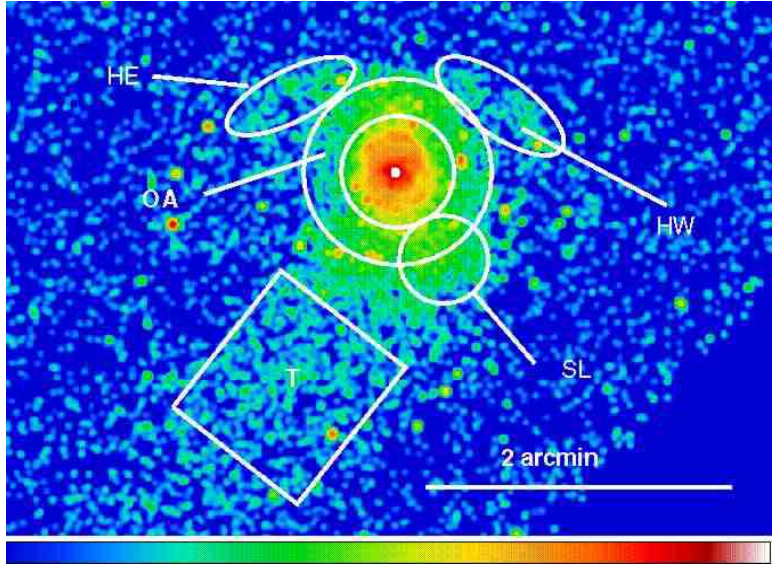}
\caption{\footnotesize{ Background subtracted, exposure corrected $0.5-2$\,keV
{\it Chandra} image of NGC~4552 overlaid with
the spectral extraction regions 
for gas at large radii (OA), in the horns (HE, HW), 
in the southern lump (SL), and
in the  tail (T), defined in Table  
\protect\ref{tab:allspectra}. 
The image has been smoothed with a Gaussian kernel with 
$\sigma = 1''$. 
}}
\label{fig:specimage}
\end{center}
\end{figure}

Spectral extraction regions were chosen to isolate, as much as 
possible given our limited photon statistics,  
emission from the individual features. 
The geometries of these spectral extraction regions are listed in 
Table \ref{tab:specregs} and are shown in 
Figure \ref{fig:specimage}.

These regions are HE and HW for the eastern and western 'horns' 
extending from the northern surface brightness edge to either side 
of the galaxy, OA for galaxy emission at large radii but inside the 
northern edge, 
SL for the region ('lump') of enhanced emission in the 
southwestern quadrant of NGC~4552, and T for the tail of emission
extending from the galaxy to the south. 
To determine source existence, one compares the source plus background
counts in a region to that expected from the background alone. 
We also can determine how well the flux can be measured for each feature.  
We compute both these statistics for
regions HE, HW, SL, and T. The background
regions we used to compute these measures are 
listed in Table \ref{tab:bgsigregs}, and the respective areas and
observed counts are given in Table \ref{tab:fluxsigdat}. We note that 
the background region (SL\_BG) used for region SL, that is embedded in the
galaxy, was a  neighboring region that included 
the surrounding galaxy emission. Neighboring Virgo Cluster emission 
regions were used as local backgrounds to compute these statistics 
for features HE, HW, and T, since those regions extend outward 
from NGC~4552 into the Virgo Cluster emission. For each region we find 
that the cumulative Poisson probability that the observed or greater 
number of counts could have arisen from a fluctuation of the 
expected background is $\lesssim 10^{-10}$. 
Thus the detection of these features is highly significant. 
The precision to which the flux can be measured for each feature 
is given in Table \ref{tab:fluxsigdat}, as the net source counts 
and the total uncertainty in each measurement.

For the analysis below, since the spectra for gas in NGC~4552 and 
in the X-ray tail are soft, we restrict the spectral fit energy range
for the tail (T), horns (HE+HW) and galaxy regions outside the rings 
to $0.3-2$\,keV, where the source count rates are above background.
We initially separated the outer annular region (OA) for NGC~4552 
into two separate rectangular regions, one just inside the northern 
leading edge (NG)  and one to the south near the beginning of the 
tail (SG), to look for temperature differences between these two 
regions. However, since we found no statistically significant difference 
between the spectral fits for these regions 
(see Table \ref{tab:allspectra}), we combined them into the annular 
region (OA) shown, in order to improve statistics and sharpen our 
measurement of the temperature and abundance for galaxy gas outside 
the central region. 

\subsubsection{Virgo North (VN) Background Region}

We chose a $6'.7 \times 2'.5$ rectangular region (VN) centered $3'$ 
north of NGC~4552, comprising approximately $25\%$ of the area of 
the S3 chip, to determine the spectrum of the Virgo Cluster gas in the
vicinity of 
NGC~4552. We fit the spectrum for the Virgo ICM (region VN) over the 
full $0.3-7$\,keV energy band, where the S3
detector has good efficiency.
We fix the hydrogen absorbing column at 
$n_{\rm H} = 2.59 \times 10^{20}$\cms, the Galactic value 
(Dickey \& Lockman 1990) and allow the temperature, abundance and
normalization to vary. We find the temperature of the
Virgo Cluster ICM to be $2.2^{+0.7}_{-0.4}$\,keV with metal abundance
$A=0.1^{+0.2}_{-0.1}$. We then allowed the absorbing column to vary
and found no suggestion for increased absorption. Our results are 
 in excellent agreement with previous 
spectral analyses of the Virgo Cluster ICM using data from 
{\it Ginga} (Takano \etal 1989), 
{\it ROSAT} (B\"{o}hringer \etal 1994), and {\it ASCA} (Ohashi \etal 1998, 
Shibata \etal 2001), that find, with Galactic hydrogen absorption,  
 temperatures $\sim 2$\,keV and metal
abundances of $\sim 0.1 - 0.3\,\Zs$, at projected distances 
($\sim 300$\,kpc from M87) comparable to that of NGC~4552. 
We note that the temperature and X-ray emissivity of the Virgo ICM are 
relatively insensitive to the metal abundance over the abundance 
range ($A < 0.3\Zs$) determined by our fit. 
If we allow the abundance to vary over this range, we
find variations in the Virgo ICM temperature  
of $\lesssim 0.4$\,keV, within the $90\%$ confidence level for
the canonical fit, and variations in the inferred 
$0.5-2$\,keV X-ray emissivity for the cluster gas of $\sim 10\%$.

Region VN also was used  as a local background region 
for galaxy source regions  at large radii (NG, SG, OA, SL),  the 
horns (HW+HE) and  the tail (T), to remove contamination from Virgo 
Cluster emission. For spectral fitting, counts were grouped both by 
using predefined bins resulting in channels of approximately 
logarithmic width, and by using
a minimum $20$ counts per bin. Both methods gave consistent spectral 
model fits. The spectra for all of the above regions
are well represented by single-temperature APEC or VAPEC 
thermal plasma models (Smith \etal 2001) corrected for absorption
using Wisconsin photo-electric cross-sections (Morrison \& McCammon
1983). Our results for these regions are summarized in 
Table \ref{tab:allspectra}.

We checked that our results were not sensitive to the position of the  
local background region (VN) on the detector by recomputing the 
spectral fits using a second $100''$ circular 
region $4'$ east of NGC~4552 (VE in Table \ref{tab:specregs}) and found no 
significant differences in the fitted parameters.
We also fit the spectra to these regions using blank sky backgrounds 
(as for VN) and a two component APEC + APEC model. We fixed the temperature 
and abundance of one APEC component at our best fit values for the 
Virgo Cluster emission ($kT=2.2$\,keV and $A=0.1\,\Zs$, see Table 
\ref{tab:allspectra}), while allowing the temperature and abundance of
the second APEC component and both component normalizations to vary, and 
found consistent results. Thus our results are not biased by our 
choice of VN as the local background. Since the 
models agree within their $90\%$ CL, we present only those results (using
background region VN) for the tail and galaxy emission at large radii 
in Table \ref{tab:allspectra}. 

\subsubsection{Outer Annulus (OA) of ISM}

For fixed Galactic absorption, we find the temperature and 
abundance of the outer galaxy gas in region OA to be 
$kT=0.43^{+0.03}_{-0.02}$\,keV and $A = 0.4^{+0.2}_{-0.1}\,\Zs$, 
a metallicity marginally higher than that for the Virgo ICM, 
but with large uncertainty. To check the stability 
of the fit, we allowed the absorbing column to vary and found no 
suggestion for increased absorption above the Galactic value.
To assess the effect of any residual unresolved X-ray 
binaries on our results, we also fit the spectrum for region OA, 
galaxy emission within the northern edge, over the $0.3-5$\,keV energy
range with an absorbed two component model consisting of an APEC 
component for emission from the diffuse gas and a $5$\,keV bremsstrahlung
component to model unresolved X-ray binaries (Kraft \etal 2001). We 
fix the hydrogen absorbing column at its Galactic value, while 
allowing the temperature and abundance in the APEC component and 
both component normalizations to vary. We find that  
the fitted temperature ($0.42^{+0.03}_{-0.02}$\,keV) and abundance 
($0.4^{+0.2}_{-0.1}\,\Zs$) for the thermal component in this two 
component model agree, within their $90\%$ confidence limits,    
with the single APEC model results listed in Table
\ref{tab:allspectra}. The bremsstrahlung (unresolved binary) 
component contributes only $\sim 5-6\%$ to the $0.5-2$\,keV luminosity of
region OA. This is consistent with theoretical expectations for the 
contribution of X-ray binaries to the flux in region OA found by 
integrating the average X-ray luminosity function for LMXB's 
(Gilfanov, 2004) below our observed $0.5-8$~keV point source detection 
threshold ($1.5 \times 10^{-15}$\ergscm) in that region.
Thus, the effects of residual unresolved X-ray binaries on the measured
properties of the diffuse gas in region OA 
do not bias the single temperature APEC model results for the
properties of the galaxy gas which are important to our analysis.
We find that the temperature of the ISM on the galaxy side of the 
(leading) northern edge ($0.43$\,keV) is a factor $\sim 5$ lower than the
temperature of the gas in the ambient Virgo ICM ($\sim 2.2$\,keV). 
As shown in 
Figure \ref{fig:image}, this temperature drop corresponds to 
a sharp increase in surface brightness (and thus gas density as 
quantified in \S\ref{sec:sboutprofile}). This is 
 consistent with the interpretation of the northern edge as a 
cold front, the leading edge of the galaxy ISM as
it undergoes ram pressure stripping, in broad agreement with 
expectations from simulations (e.g. see Stevens \etal 1999, Figure 2, 
model 1b; Acreman \etal 2003, Figures 10a \& c and 11b \& d, $1.6$\,Gyr 
time slice).

\subsubsection{ Southern `Lump' (SL) Spectrum}

Figures \ref{fig:image} and \ref{fig:panda} show a bright region 
(lump) of emission 
in the outer southwest quadrant of the galaxy. We use a $17''$ circular
region centered at ($\alpha =12^h35^m38.5^s$,
$\delta = +12^\circ32'48''.2$) surrounding the lump   
(region SL in Figure \ref{fig:specimage})
to isolate the spectrum for this region. Although we find, using  
an APEC model with Galactic absorption,  a
temperature $kT =0.40^{+0.03}_{-0.02}$\,keV and abundance $A > 0.5$ 
formally consistent with the properties of galaxy gas elsewhere 
outside the central rings in NGC~4552, the 
$\chi^2/{\rm dof} = 39/26$ is large.
The fit is improved 
($\chi^2/{\rm dof} = 29/26$) by using an absorbed VAPEC model 
with a higher iron abundance, while fixing all 
other abundances at $A = 0.4\,\Zs$, the mean abundance  
for the galaxy ISM in region OA. The hydrogen absorbing column remained fixed
at the Galactic value. We include this fit in Table \ref{tab:allspectra}.
In this VAPEC model, the temperature of the gas in the southern lump is 
$kT = 0.38 \pm 0.03$\,keV,  consistent with previous results for gas at 
large radii, but the iron abundance  is increased to
near solar values ($A_{\rm Fe} = 0.8 \pm 0.2\,\Zs$). This may suggest 
that the southwest quadrant of the
galaxy has either experienced a recent episode of star formation 
itself or has had metals transported from star forming activity near 
the galaxy center (expected for a transition type nucleus) to 
larger radii in the galaxy as a result of outflows.

\subsubsection{East and West Horns (HE, HW)}

The regions HE and HW for the horn-like 
features are faint, with region HE and HW containing only 
$177 \pm 13$ and $155 \pm 12$ net source counts, respectively, in the 
$0.3-2$\,keV energy band.
In order to improve statistics, we combine these regions
and fit their average spectrum (HE+HW  in Table
\ref{tab:allspectra}). We restrict our models to those with 
three or fewer free parameters. Neither an absorbed bremsstrahlung model nor 
absorbed power law model, each with all parameters free to vary, can 
fully describe the data, giving $\chi^2/{\rm dof}$ of $54/15$ and $61/15$, 
respectively. A thermal APEC model with Galactic absorption provides a
better fit to the data with temperature 
$kT =0.46^{+0.08}_{-0.06}$\,keV and abundance 
$A = 0.2 \pm 0.1\,\Zs$ ($\chi^2/{\rm dof} = 25.7/15$). The
probability of obtaining this value of $\chi^2$ or higher, given 
$15$ degrees of freedom,  is $0.042$ ($2.1\,\sigma$ confidence). 

The $\chi^2$ for the model fit to the combined spectrum for the horns
might suggest that these two regions are not homogeneous in their 
spectral properties. We investigate this possibility by fitting the 
spectrum for each horn individually. Using absorbed APEC thermal 
plasma models with fixed abundance and Galactic absorption, we find 
the temperature of gas in the eastern horn HE     
to be $kT = 0.39^{+0.08}_{-0.06}$\,keV ($\chi^2/{\rm dof} = 4.4/8$) and 
in the western horn HW to be $kT = 0.54 \pm 0.08$\,keV 
($\chi^2/{\rm dof} = 4/4$) for $0.2\,\Zs$ abundance, taken from the
fit to the combined horn regions HE+HW (see Table \ref{tab:allspectra}). 
These temperatures vary by $\lesssim 5\%$ when the abundance is varied 
between $0.2$ to $0.4\,\Zs$ (see Table \ref{tab:metalscool}). 
Although suggestive, the fitted temperatures agree within their $90\%$
confidence limits. Thus with these data, we can not conclude  
that the temperatures of the eastern and western horns are significantly 
different. However, if such a temperature difference were to be
confirmed in a deeper exposure, it would not be surprising. The 
horns are spatially separated and most likely correspond to 
different stripping events. Temperature differences  between gas in 
the two horns could arise from hydrodynamical effects caused by 
local temperature and pressure gradients in the different turbulent 
eddies responsible for the stripping process in each region.

The main conclusion that can be drawn from the current data is 
that the temperature of the gas in the horns is 
cool, similar to that measured throughout the outer regions of
NGC~4552 (see Table \ref{tab:allspectra}. These temperatures 
agree within their $90\%$ confidence 
limits with the temperature ($kT= 0.43^{+0.03}_{-0.02}$\,keV) 
for gas in the outer radii of NGC~4552 (region OA). In contrast,  
they are $4$ times lower than the $2.2$\,keV temperature of 
the surrounding Virgo Cluster ICM. This is the 
key property of the gas in the horns that identifies it 
as galaxy (and not cluster) gas, likely in the process of 
being stripped. These results are consistent with 
simulations of ram-pressure stripping
(see, e.g. Stevens \etal 1999, Figure 2, model 1b), where 
irregular filaments due to the onset of hydrodynamical instabilities
during stripping are cooler and denser than the surrounding ICM.

\subsection{Gas Densities from Fitting the Surface Brightness Profile}
\label{sec:sboutprofile}
\begin{figure}[t]
\begin{center}
\includegraphics[height=2.273in,width=3in]{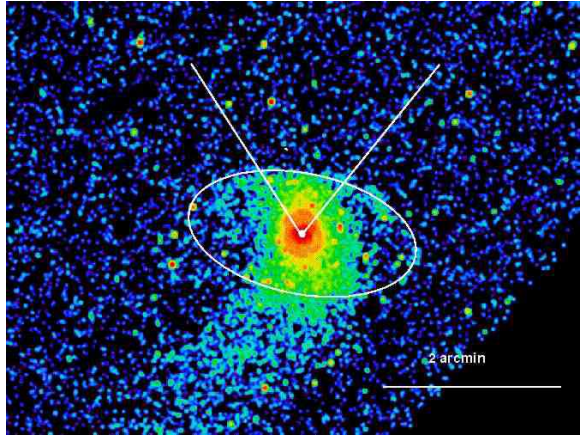}
\caption{\footnotesize{ Background subtracted, exposure corrected $0.5-2$\,keV
{\it Chandra} image of NGC~4552 overlaid with
the angular sector and bounding ellipse used to extract the surface
brightness profile shown in the left panel of Figure 
\protect\ref{fig:sbprof}. 
The image has been smoothed with a Gaussian kernel with 
$\sigma = 1''$. 
}}
\label{fig:edgeimage}
\end{center}
\end{figure}


In Figure \ref{fig:edgeimage} we show the angular sector used to
construct the $0.5-2$\,keV surface brightness profile to the north 
across the flattened surface brightness edge 
between galaxy gas and the Virgo Cluster ICM, as a function 
of the mean distance from the center of NGC~4552.
 The profile is
constructed from elliptical annuli constrained to lie within the 
angular sector centered on the galaxy
nucleus and extending from $57^\circ$ to $129^\circ$ 
measured clockwise from east, that was chosen   
to exclude the horns of emission extending
southward from each side of the northern leading edge.  
The elliptical annuli are 
concentric to a `bounding' ellipse, with semi-major (-minor) axes  
of $78''$ ($40''.5$) and
position angle of $349^\circ$, 
that traces  the shape of the northern surface brightness edge
within the angular sector. The radial extent of each elliptical annulus
varies logarithmically inward and outward from the bounding ellipse 
with logarithmic step size $1.1$. 

\begin{figure}[t]
\begin{center}
\epsscale{0.5}
\epsfig{file=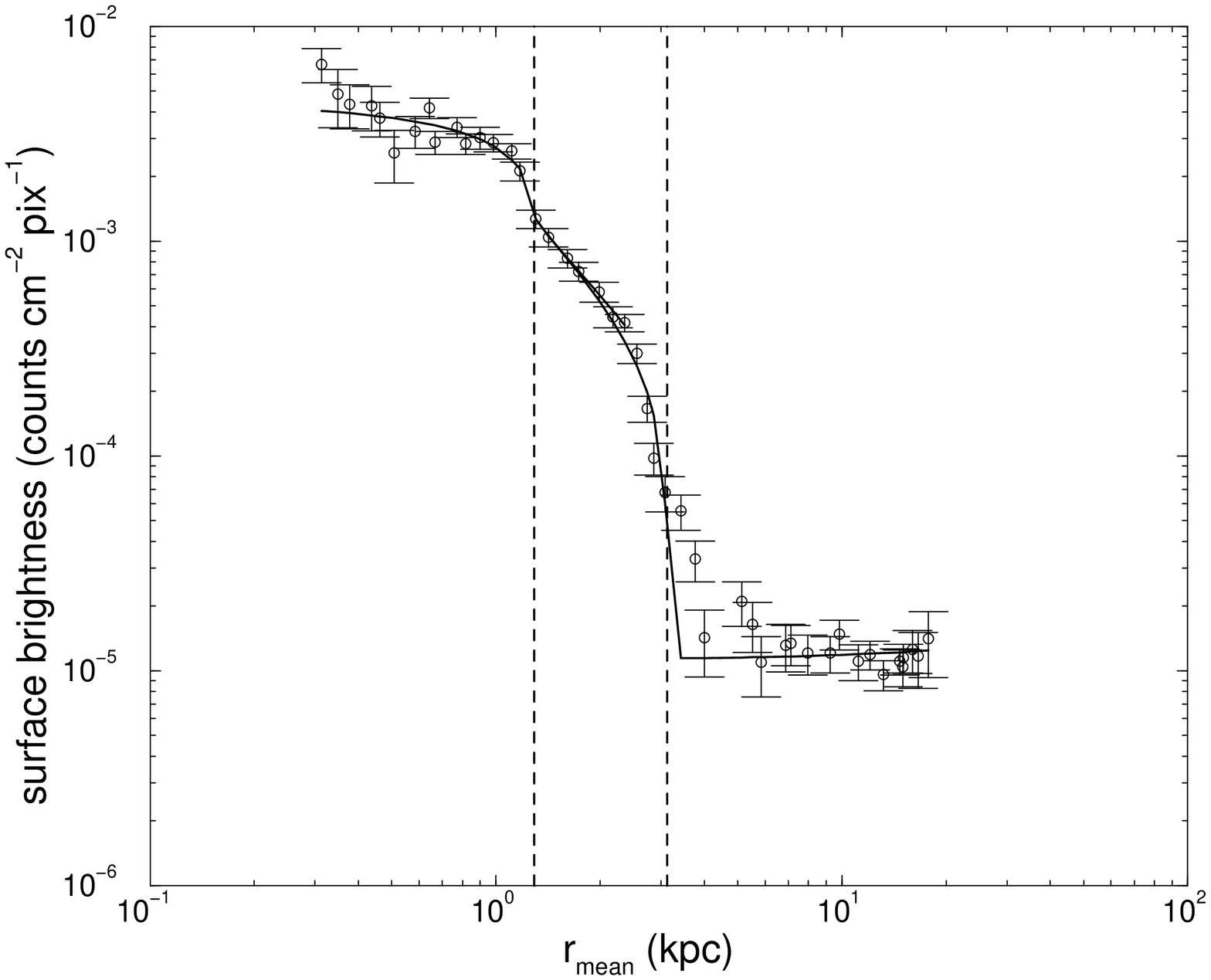,height=2.75in,width=2.75in,angle=270}
\hspace{0.1cm}
\epsfig{file=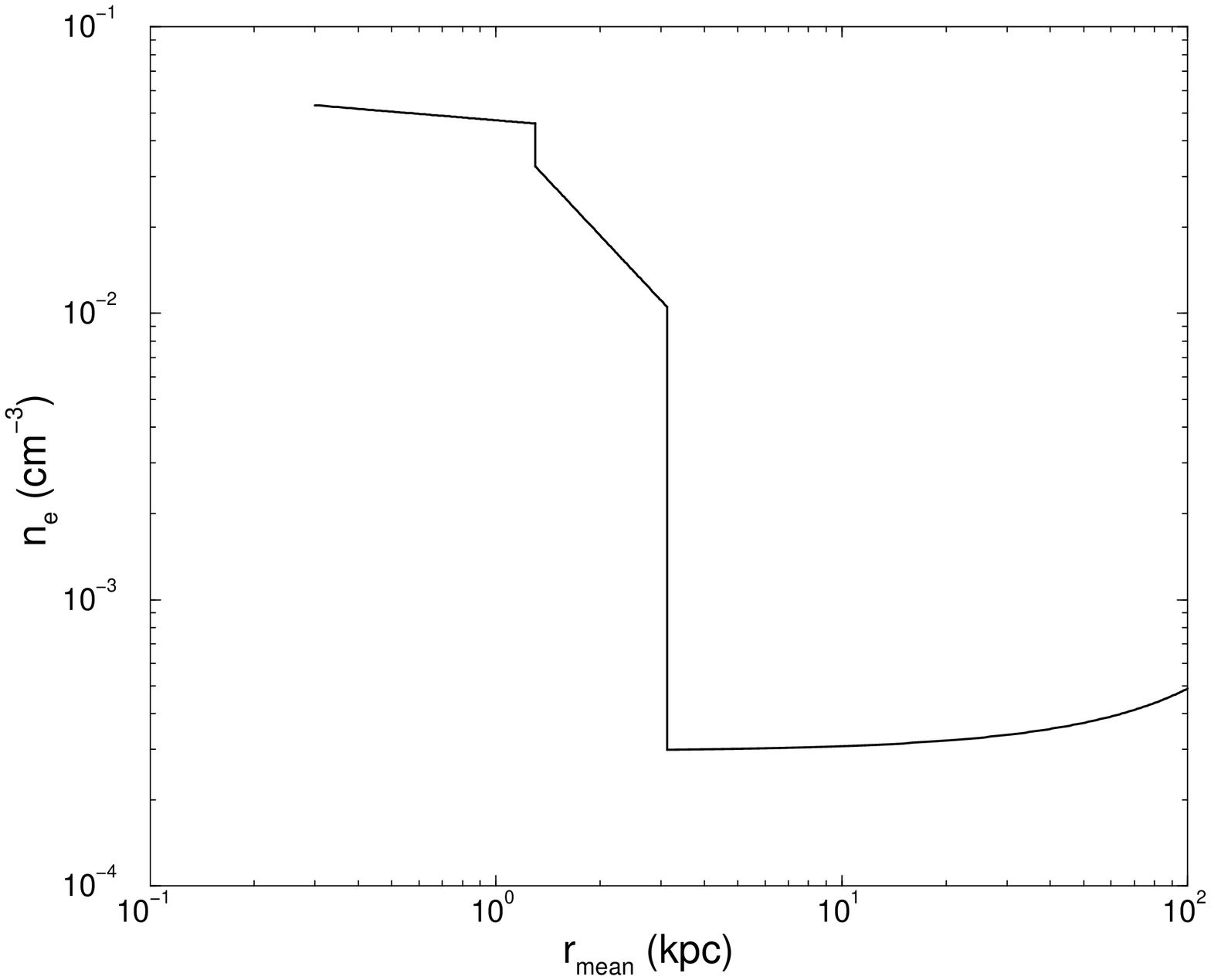,height=2.75in,width=2.75in,angle=270}
\caption{\footnotesize{(left) The $0.5-2$\,keV surface brightness profile 
as a function of distance from the center of NGC~4552 toward the north
across the galaxy's leading edge. 
The vertical dashed lines denote $r_1=1.3$\,kpc, the position of the 
outer radius (rim) of the northern ring (see Paper II), and $r_2=3.1$\,kpc, 
the outer leading edge of the cold front.  
The solid line denotes the 
model results for the surface brightness for power law density 
models ($n_e \propto r^{-\alpha}$ within the galaxy with
$\alpha = 0.1^{+0.25}_{-0.1}$ for $r <1.3$\,kpc 
(inside the rim) and $\alpha = 1.3^{+0.2}_{-0.2}$
for $1.3 \le r < 3.1$\,kpc (between the rim and the leading edge of
the cold front).  
For $r \ge 3.1$\,kpc the Virgo ICM is modeled
by a $\beta$ model centered on M87 with index $\beta = 0.47$ and 
core radius $r_c = 2'.7$ (Schindler \etal 1999).
(right) The electron density as a
function of radius from the center of NGC~4552 for the above model.
}}  
\label{fig:sbprof}
\end{center}
\end{figure} 

In the left panel of Figure \ref{fig:sbprof}, we show the 
$0.5-2$\,keV surface brightness
profile as a function of the mean distance from the center of NGC~4552
to the north across the flattened surface brightness edge 
between galaxy gas and the Virgo Cluster ICM.
Two sharp surface brightness discontinuities (denoted by vertical
dashed lines) are evident in Figure \ref{fig:sbprof}: the  outer, `leading' 
edge is at the northern interface between the galaxy cold front 
and Virgo Cluster gas,
and the inner discontinuity is at the position of the rim of the 
bright northern ring. We analyse the leading edge of the cold front 
in this paper and the central rings in Paper II.

The observed surface brightness profile is proportional to the 
product of the X-ray emissivity and the square of the electron 
density, integrated along the line of sight. 
By fitting the surface brightness profile, we are 
able to determine the shape of the density distribution, the location
of the edge, and the ratio of electron densities on either side of 
the edge, averaged over the respective profile bins. 
If the density is slowly 
varying over the profile bin, as is the case for a cold front, this 
measured density ratio is a good measure of the physical density 
discontinuity across the edge.

As was done for other cold fronts (Markevitch \etal 2000; Vikhlinin
\etal 2001; Machacek \etal 2005a), we assume a spherically symmetric 
power law distribution of the form 
\begin{equation}
 n_{\rm e} = n_{e2}\Bigl (\frac{r}{r_2} \Bigr )^{-\alpha_2}
\label{eq:galouter}
\end{equation}
for the electron density in NGC~4552 inside the leading
edge of the cold front (but outside the central rings), 
where $n_{\rm e2}$ and $\alpha_2$ are the
normalization and power law index for the electron 
density of NGC~4552 in the cold front, respectively. 
The radial distance $r$ and location $r_2$ of the leading 
surface brightness discontinuity are measured from the optical 
center of NGC~4552, i.e. the center of the bounding ellipse. 

We model the electron density in the Virgo Cluster ICM  
(within subcluster A) using a 
spherically symmetric, isothermal $\beta$ model centered on M87 
with $\beta = 0.47$,  and core radius $r_c = 2'.7$
taken from a fit by Schindler \etal (1999) to 
{\it ROSAT} All-Sky Survey data of the Virgo ICM 
out to a radius of $250'$ from M87. 
We normalize the cluster 
$\beta$ model using the surface brightness $\sim 8$\,kpc away from
NGC~4552, in the undisturbed Virgo ICM (cluster free stream region).
We find the central electron density (density $\beta$ model normalization) 
for the Virgo ICM to be $n_0 = 0.030 \pm 0.003$\cmc, in agreement 
with previous results (Schindler \etal 1999).

However, to model the electron 
density, at the interface between galaxy gas in the cold front and the 
Virgo ICM, requires additional assumptions about the 
three-dimensional geometry of NGC~4552 relative to the cluster
center. We assume that the distance of M87 is representative of the 
cluster center, at least for subcluster A containing both M87 and NGC~4552.
Measurements of distance moduli $DM$ for NGC~4552 
($DM = 30.93 \pm 0.14$) and M87 ($DM = 31.03 \pm 0.16$), using I-band 
surface brightness fluctuations (Tonry \etal 2001), agree
within uncertainties, suggesting the two galaxies are at 
comparable distances. We thus assume that  
the $71.'6$ projected distance between NGC~4552 and M87 is 
representative of their true physical separation. The  
electron density of the Virgo ICM, just outside the 
leading edge,  is then $n_{ec} = 3.0 \pm 0.3 \times 10^{-4}$\cmc. 

We then use a  multivariate 
minimization algorithm to fit the surface brightness profile across
the leading edge, allowing
the location of the edge ($r_2$) , the power law index 
of the electron density inside the edge ($\alpha_2$), and the 
discontinuity ($J$) to be free parameters. The discontinuity ($J$)
is given by 
\begin{equation} 
 J = \Bigl (\frac{\Lambda_{\rm in} n_{\rm in}^2}{\Lambda_{\rm out} n_{\rm out}^2} \Bigr )^{1/2}, 
\label{eq:jump}
\end{equation}
where $\Lambda_{\rm in}$($ n_{\rm in}$) and $\Lambda_{\rm out}$($ n_{\rm out}$) are the
X-ray emissivity (electron density) inside ($r \le r_2$) and 
outside ($r \ge  r_2$) the leading edge of the cold front.
We find the best fit position of the leading, northern edge (cold
front) to be $r_2 = 3.12^{+0.01}_{-0.02}$\,kpc from the galaxy center,
the slope of the electron density
inside the cold front to be $\alpha_2 = 1.3 \pm 0.2$,  
and the discontinuity $J$ (given by equation \ref{eq:jump}) to be   
$54 \pm 4$. The uncertainties on the discontinuity reflect the
$90\%$ CL uncertainties in the slope found for $\alpha_2$. 

While the surface brightness discontinuity $J$ depends only on the 
slope of the 
power law distribution given by the fit to the surface brightness profile,  
we see from equation \ref{eq:jump} that the ratio of the electron 
densities inside and outside the leading edge is sensitive, through 
the X-ray emissivity, to the spectral properties of the gas.
In Table \ref{tab:metalscool} we show the dependence of the spectral
fit parameters on the galaxy abundances for all the galaxy regions 
of interest and list the parameters for the nearby Virgo Cluster ICM
for comparison. As this table shows, the temperature 
found for gas in each region is insensitive to the abundance
over the range of interest ($\sim 0.2 - 1\,\Zs$). 
On the other hand, since metal lines dominate the cooling 
function for temperatures $\lesssim 1$\,keV (Tucker \& Gould 1966; 
Raymond \& Smith 1977; Smith \etal 2001), the spectral model 
normalization, and thus X-ray emissivity, vary by 
a factor of $\sim 4$ 
over the same range. As shown for the infalling elliptical
galaxy NGC~1404 in the Fornax Cluster  
(Machacek \etal 2005a), this may introduce significant uncertainties 
in modeling the electron density in the galaxy from the surface brightness 
discontinuity between galaxy and cluster gas.

We use  $0.5-2$\,keV X-ray emissivities given in Table \ref{tab:metalscool}
derived from our spectral fits, the cluster ICM density outside the 
cold front ($n_{ec} = 3.0 \times 10^{-4}$\cmc) 
determined from the cluster ICM $\beta$-model, and the measured
discontinuity ($54 \pm 4$) in equation \ref{eq:jump} to infer  
the electron density $n_{e2}$ inside the cold front. 
In Table \ref{tab:properties} we show the effect of the uncertainty 
in NGC~4552's abundance (and thus emissivity) on the ratio of electron 
densities inside and outside the leading edge (cold front) and the 
propagation of that uncertainty through the analysis of the galaxy
motion that follows. 
Since we found no strong metallicity gradients within the galaxy
(see Table \ref{tab:allspectra}), we adopt a 
common mean metallicity for gas in the whole galaxy of $A =0.5\Zs$, 
determined from a spectral fit to the
galaxy emission within a radius of $50''$, excluding resolved point 
sources and the nucleus. From Table \ref{tab:metalscool}, the 
$0.5-2$ keV emissivity for $0.42$\,keV galaxy gas at large radii (region OA) 
with abundance $A=0.5\Zs$ is 
$\Lambda_{\rm in} = 1.0 \times 10^{-23}$\,erg\,cm$^3$\,s$^{-1}$.
We take the $0.5-2$\,keV X-ray emissivity for $A=0.1\,\Zs$  Virgo ICM 
gas (also from Table \ref{tab:metalscool}) to be 
$\Lambda_{\rm out} = 5 \times 10^{-24}$\,erg\,cm$^{3}$\,s$^{-1}$.  
The density ratio across the cold front is then found to  
be $n_{e2}/n_{ec} = 38$, and the electron density at $r_2=3.1$\,kpc, inside 
NGC~4552 in the cold front, is $n_{e2}=0.01$\cmc. 

\subsection{The Leading Edge of the Cold Front: 
  Constraining NGC~4552's Velocity}
\label{sec:nedge}

Following Vikhlinin \etal (2001), we use the gas temperature and 
density ratios across the leading edge 
in NGC~4552 to calculate the pressure ratio $p_2/p_c$ between galaxy 
gas in the cold front inside the leading edge and  undisturbed 
Virgo cluster gas in the free-stream region. We assume the galaxy gas in 
NGC~4552 just inside the edge is in pressure equilibrium with 
cluster gas at the stagnation point just outside the edge, where the 
relative velocity between NGC~4552 and the cluster gas is zero. 
Thus $p_2/p_c$ is a measure of the pressure ratio between cluster gas 
at the stagnation point and in the free-stream region, and can be 
used to determine the Mach number $M$ of NGC~4552's motion relative
to  the ICM (Landau \& Lifschitz 1959; Vikhlinin \etal 2001). 
In Table \ref{tab:properties} we summarize the dependencies of this  
velocity analysis for NGC~4552 on the metallicity of the galaxy gas 
over the abundance range $0.2 - 1.0\,\Zs$. 
It is important to note that, because 
the motion lies on the supersonic, steeply rising branch of the 
pressure ratio versus Mach number curve (Eq. $3$ in Vikhlinin \etal 2001),
the Mach number remains well constrained, 
despite the uncertainties in the density and pressure ratios caused by 
uncertainties in NGC~4552's poorly constrained metallicity.

We find that NGC~4552 is moving at Mach $M = 2.2^{+0.5}_{-0.3}$ 
through the Virgo Cluster gas.  Given the speed of sound 
in $2.2$\,keV gas of $c_s = 766$\kms, the speed  of 
NGC~4552 relative to the ICM is $v = 1680^{+390}_{-220}$\kms. 
Taking $v_r = -967 \pm 11$\kms, the relative radial velocity 
between NGC~4552 and M87 (NED, Smith \etal 2000), 
as the relative radial velocity 
 between NGC~4552 and the surrounding Virgo ICM,
we find the component of velocity in the plane of the sky to be  
$v_t = 1380^{+450}_{-290}$\kms
and the inclination angle of the motion with respect to
the plane of the sky to be $\xi = 35^{\circ}\pm 7^{\circ}$ towards 
the observer.

\subsection{The Ram-Pressure Stripped Tail}
\label{sec:tail} 

With the motion of NGC~4552 through the ICM constrained by the 
properties of the leading cold front, we can examine the properties
of the hot gas in the tail.  
From Table \ref{tab:allspectra}, we see that 
the temperature ($kT = 0.51^{+0.09}_{-0.06}$\,keV) 
of X-ray-emitting gas in the tail  
is consistent (given the large uncertainties) with
the temperature of gas in the outer region of the galaxy OA 
 ($kT=0.43^{+0.03}_{-0.02}$) 
and in the horns ($kT =0.46^{+0.08}_{-0.06}$), 
where stripping is expected to be occurring. The gas temperature in 
the tail is $\sim 4$  times lower than that of the Virgo ICM,  
proving that the tail is composed predominantly of 
ram-pressure-stripped galaxy gas rather than Virgo ICM gas concentrated
into a wake. Although poorly constrained due to our limited statistics,
the metallicity of gas in the tail 
($A \sim 0.4^{+1.1}_{-0.2}\,\Zs$) 
is also consistent with that in the galaxy. The $0.5-2$\,keV luminosity
for hot gas in the tail (from region T) is $\sim 9 \times 10^{38}$\ergs.  

We estimate the mean electron density $n_{\rm et}$ for gas in 
the tail from the XSPEC APEC spectral model 
normalization $K$, where 
\begin{equation}
K = \frac{10^{-14}n_{\rm et} n_{\rm Ht} \eta V}{4 \pi (D_A (1+z))^2}\,. 
\label{eq:xspec}
\end{equation} 
$D_A$ is the angular size distance to the source, $z$ is the
source redshift, $n_{\rm et}$ and $n_{\rm Ht}$ are the mean electron 
and hydrogen number densities, respectively, $\eta$ is the gas filling
factor, and $V$ is the volume of the tail emission region. 
All quantities are in cgs units. 
If we assume gas in the tail uniformly fills a cylindrical volume  
(corresponding to the spectral extraction region T) of radius 
$2.4$\,kpc ($31''$) and length $5.3/{\rm kpc \,cos}\xi$, where
$\xi$ is the inclination angle of the motion with respect to the plane
of the sky, we find a mean electron density for hot gas in the tail of 
$n_{\rm et} = 6 \pm 2 \times 10^{-3}\,{\rm cos}^{1/2}\xi$\cmc. 
As shown in Table \ref{tab:metalscool} and reflected by the large
uncertainties, the APEC model normalization $K$, and thus the inferred 
mean electron density, 
are  strongly dependent on the metallicity of the tail. 
Assuming the  ``best fit'' model for the tail
($A=0.4\,\Zs$, $kT = 0.51$\,keV, $\xi = 35^\circ$), 
the mean electron density 
in the tail (region T) 
is $5.4 \times 10^{-3}$\cmc.
Thus the tail is cooler and denser than the surrounding ICM, as 
expected from simulations 
(see, e.g. Stevens \etal 1999, Figure 2, Model 1b).

\begin{figure}[t]
\begin{center}
\epsscale{0.5}
\epsfig{file=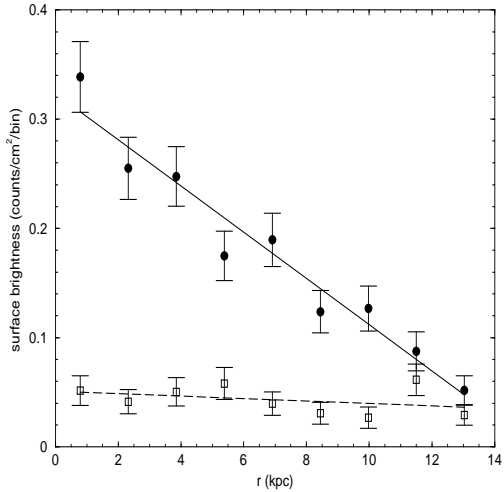,height=2.6in,width=2.6in,angle=270}
\caption{\footnotesize{ Projected surface brightness from the $0.5-2$\,keV image in 
Fig. \protect\ref{fig:image} onto the midplane of the tail, 
measured from $4$\,kpc south of NGC~4552's
center, where the tail appears to deviate from the 
hot gas halo of NGC~4552. The surface brightness is summed in 
$1.5$\,kpc bins from a rectangular region of width $5$\,kpc and 
orientation angle $226^\circ$. Open squares represent the projected
surface brightness from a parallel region of the
same dimensions in the nearby ICM. The solid 
and dashed lines represent linear regression ($y=ar+b$) fits for the 
tail and Virgo ICM with slope 
$a = -0.021 \pm 0.002 $\sbunit\,kpc$^{-1}$ and
intercept $b = 0.32 \pm 0.02$\sbunit  
for the tail, and 
$a = -0.001 \pm 0.001$\sbunit\,kpc$^{-1}$  and 
$b = 0.05 \pm 0.01$\sbunit for the nearby Virgo ICM.
}}
\label{fig:tailicmproj}
\end{center}
\end{figure}

In Figure \ref{fig:tailicmproj} we compare the 
surface brightness in a rectangular region of width 
$5$\,kpc  projected onto the tail's 
central axis in $1.5$\,kpc bins along the length of the tail
(filled circles) to the projection of the surface brightness of the 
surrounding Virgo ICM in a parallel region of the same dimensions,
orientation, and bin size (open squares).
We define the beginning of the tail (and $r=0$ of
the projection in Figure \ref{fig:tailicmproj}) at 
the point $\sim 4$\,kpc south of NGC~4552's center,  
where the tail visually appears distinct from the galaxy's   
more spherical, hot gas halo (see Figure \ref{fig:image}).
The solid and long-dashed lines denote the best linear fits to the 
projected surface brightness of the tail and Virgo ICM, respectively. 
As expected from the $\beta$-model for the Virgo ICM 
(Schindler \etal 1999), the projected surface
 brightness of the Virgo ICM at the distance of the tail from M87 
 is very flat, consistent with being constant.  In contrast the 
projected surface brightness of the tail declines 
by a factor $\sim 5$  before becoming indistinguishable 
from the surrounding Virgo ICM. Defining the length of the observed 
tail, in the plane of the sky, as the distance from where the tail 
appears distinct from NGC~4552 to where it fades into the Virgo 
ICM, we estimate  this length to  
be $\sim 11$\,kpc (from Figure \ref{fig:tailicmproj}), 
corresponding to a physical length (given
$\xi=35^\circ$) of $\sim 13$\,kpc. Using our derived  mean density of 
$\sim 5 \times 10^{-3}$\cmc and assuming uniform filling of a 
cylindrical volume of radius $2.4$\,kpc and physical length $13$\,kpc,
we estimate the gas mass in the observed tail
to be  $M_{tail} \sim 3 \times 10^7\Ms$.

We find the mean thermal pressure for gas in the tail to be  
$\sim 6 \pm 3 \times 10^{-3}$\,keV\,\cmc, where the errors reflect the 
$90\%$ CL combined uncertainties due to the measured tail gas temperature 
and metallicity and NGC~4552's inclination angle $\xi$.  
We estimate the mean 
pressure of the surrounding ICM using the Virgo ICM $\beta$-model 
evaluated at the midpoint of the cylindrical emission region (T).
Since the perpendicular ($\xi$ dependent) component of the distance 
between the midpoint of region T and the Virgo Cluster center 
(taken to be M87) is small 
($\sim 0.'9 - 1'.4$ for $\xi \sim 30 -40^\circ$) compared 
to the transverse component ($72'$), the $\xi$ dependence of the 
cluster gas electron density can be ignored. 
We find the mean thermal pressure for
undisturbed Virgo Cluster gas at the location of the 
tail to be $\sim 1.2^{+0.5}_{-0.3} \times 10^{-3}$\,keV\cmc. 
Although the uncertainties 
are large, these data suggest that the thermal pressure of the
undisturbed ICM alone may not be sufficient to establish pressure
equilibrium with the stripped gas in the tail. This is in agreement 
with simulations that find the stripped ISM may not be in pressure 
equilibrium with the ambient medium, but that gas in the tail 
expands while it is removed (Toniazzo \& Schindler 2001; Heinz \etal 2003). 

If we assume 
the gas, once heated and stripped from the galaxy, 
 is free to expand adiabatically at
approximately the tail sound speed 
($c_t \sim 370^{+30}_{-20}$\kms for $0.51^{+0.09}_{-0.06}$\,keV gas), 
we can use the initial tail radius $r_{\rm init}$ to estimate 
the time $t_{\rm eq}$ needed for gas in the tail  
to reach pressure equilibrium with the undisturbed ICM 
(Machacek \etal 2005b), i.e. 
\begin{equation}
 t_{\rm eq} = 
  \frac{r_{\rm init}}{c_t} \Bigl (\frac{p_t}{p_c} \Bigr )^{0.3}
\end{equation}
for a monatomic ideal gas with adiabatic index $\gamma = 5/3$, 
where $p_t$ and $p_c$ are the initial and final thermal 
pressures in the tail. Assuming thermal pressures dominate, 
the final tail pressure $p_c$ is approximately the pressure 
of the surrounding ICM.
Using the best APEC model fit parameters for the tail gas, 
$\xi =35^\circ$, and initial radius $\sim 2.4$\,kpc (from region T),
we find the time for the tail to expand to equilibrium to be $\sim 10$\,Myr.
Given NGC~4552's transverse velocity of $\sim 1400$\,kms, 
gas at the end of the observable tail ($\sim 11$\,kpc downstream 
in the plane of the sky) has had $\sim 8$\,Myr more time to expand than 
gas now  found  near the beginning of the tail, and so the southern 
end of the tail may be nearly in pressure equilibrium with the ICM.

As the gas density in the tail decreases due to adiabatic expansion, 
we would expect the tail to broaden and fade. From conservation of entropy,  
the mean density and temperature of the
tail gas at pressure equilibrium would be $n_{et} \sim 2 \times 10^{-3}$\cmc
and $kT \sim 0.3$\,keV, with an expected corresponding decrease in 
surface brightness by a factor $\sim 8$. Although our statistics are 
too limited to directly observe a temperature decline in the tail, 
the observed rapid fading of the tail, by a 
factor $\gtrsim 5$ from 
Figure \ref{fig:tailicmproj}, is broadly consistent with that expected
from  adiabatic expansion.

\subsection{In Search of the Bow Shock}
\label{sec:bowshock}

For supersonic motion one would expect a bow shock to form in 
the Virgo ICM in front of the galaxy. However, the bow shock is
likely to be the least visible of the three features (cold front, tail, 
bow shock) expected from supersonic ram pressure stripping (Stevens
\etal 1999). The primary reason for this is that the shock front is 
narrow and the observed properties are very sensitive to
projection effects. While the Rankine-Hugonoit conditions (\S\,$85$ 
Landau \& Lifshitz 1959) predict a peak density and temperature jump 
of $2.5$ and $2.3$ at the shock front for Mach $2.2$ motion through 
the cluster gas, the observed temperature and density jumps will be much 
lower, primarily due to integration along the line of sight 
that crosses 
mostly through unshocked cluster gas, and to projection 
effects caused by the fact that the motion of NGC~4552 is not in the 
plane of the sky, but at an inclination angle of $\sim 35^\circ$ out 
of the plane towards the observer. 

Following Vikhlinin \etal (2001) we estimate the approximate 
location of the bow shock in front 
of NGC~4552 from the Mach number and the location
($x_{sb},y_{sb}$) of the body sonic point, i.e. the point on the body 
where the flow velocity equals the sound speed of the cluster gas. 
$x_{sb}$ is the coordinate of the body sonic point measured along the 
axis of symmetry and $y_{sb}$ is the radial  distance of the body
sonic point from the axis of symmetry (see Vikhlinin \etal 2001, Figure 9). 
As Vikhlinin \etal (2001) describe, we derive the shock position using
Moeckel's (1949)\footnote{Approximate Method for 
Predicting Form and Location of Shock Waves, NACA Techical Note 1921. 
Available online 
at http://naca.larc.nasa.gov/reports/1949/naca-tn-1921.} Figure 7, that
shows the ratio of the shock detachment distance $L = x_0 - x_{sb}$, 
where $x_0$ is the location of the bow shock along the symmetry axis, 
versus the radial coordinate of the body sonic point $y_{sb}$ as a 
function of the Mach number. For a Mach $2.2$ shock,  
 $L/y_{sb} \approx 0.8$. Thus if the body sonic
point is known, the calculation of the shock position is well
determined. If the body being stripped has a well defined shoulder,
as in A3667 (Vikhlinin \etal 2001), the body sonic point is at the 
location of the shoulder. However, in NGC~4552 
(see Figure \ref{fig:image}), the `horns' make the precise location
of the shoulder difficult to discern. Instead we use Figure 9 of 
Moeckel (1949)  to find the body sonic point for Mach $2.2$ motion as 
that point on the body surface where the angle, $\theta_d$, of the
tangent line from the symmetry axis to the body surface 
is  $\theta_d \approx 43^\circ$. 
Again the precise shape of the body surface is uncertain due to the
irregular `horns'. We estimate the `body surface' of the galaxy by a 
sphere centered on the galaxy nucleus with a radius of $3.1$\,kpc 
(the radial distance to the leading edge). This approximately
spherical shape is consistent with the distribution of optical light 
and broadly consistent with the galaxy gas excluding the horns and the
tail. We 
then find $y_{sb} \approx 2.5$\,kpc, the shock detachment distance 
$L \approx 2$, and the distance $x_b - x_{sb}$, between the position 
of the leading edge $x_b$ and the body sonic point along the symmetry axis,  
 of $\approx 1$\,kpc. Rewriting 
the shock detachment distance $L= x_0-x_b + x_b - x_{sb}$, we 
expect the bow shock to intersect the symmetry axis at 
$\approx 1$\,kpc outside the leading edge or equivalently at 
$r \approx 4$\,kpc from NGC~4552's center. It is
interesting to note that the surface brightness outside the leading 
edge, shown in Figure \ref{fig:sbprof}, does increase by a factor 
$\sim 2$ between $r \sim 4-6$\,kpc, broadly consistent with this 
picture. While suggestive, a much deeper observation is needed to 
establish whether or not this rise is actually evidence for the 
expected bow shock.

The density and thus surface brightness are also expected to rise in 
the stagnation region as the gas behind the shock slows and is 
adiabatically compressed in front of the leading edge 
(Vikhlinin \etal 2001). Again a 
rise in the surface brightness is observed in 
Figure \ref{fig:sbprof} in front of the northern edge 
($r \sim 3.1$kpc). However, this observed rise in surface 
brightness also could partly be due to galaxy gas being stripped 
by instabilities to form the horns. The expected temperature rise 
in the gas behind the shock front, as seen, e.g., in simulation slices 
 by Stevens \etal (1999), would not be observable in our data due to
dilution by unshocked gas projected along the line of sight and our limited
statistics over the small region (Mach angle $\sim 28^\circ$) between the 
shock front and the galaxy.

\section{CONCLUSIONS}
\label{sec:conclude}

In this work we analysed a $54.4$\,ks {\it Chandra} observation 
of the elliptical galaxy NGC~4552 in the Virgo Cluster and found 
X-ray evidence for ram-pressure stripping in the outer regions 
of the galaxy, due to its supersonic motion through the Virgo ICM.  
We find the following:
\begin{enumerate}
\item{The $0.5-2$\,keV X-ray emission shows the classic features of a 
cold front and ram-pressure stripping as a gas rich galaxy moves 
supersonically within a rich cluster. 
We see:  
 (1) a sharp, flattened leading surface brightness edge located 
$3.1$\,kpc from the center of NGC~4552 due to 
the ram-pressure of the Virgo ICM, (2) two horns of emission extending 
$3-4$\,kpc to either side of the edge that are composed of  galaxy gas 
in the process of being  stripped due to the onset of Kelvin-Helmholtz 
instabilities, and (3) a tail of emission extending 
$\sim 10$\,kpc behind NGC~4552 opposite 
the leading edge. 
}

\item{Galaxy gas inside the 
leading edge is cool ($kT = 0.43$\,keV)
compared to the surrounding $2.2$\,keV Virgo ICM. The surface
brightness distribution inside the leading edge is well fit by a 
power law density distribution $n_2 = 0.01 (r/r_2)^{-1.3}$\cmc 
(for $A \sim 0.5\,\Zs$), where $r_2 = 3.1$\,kpc is the position 
of the edge. The density ratio between gas inside the cold front and 
the surrounding Virgo ICM, found 
from fitting the surface brightness profile, is
$n_{e2}/n_{ec}=38^{+17}_{-10}$ for galaxy ISM metal abundances from 
$0.2$ to $1.0\,\Zs$.
}

\item{Assuming NGC~4552 and M87 both lie in the plane of the sky,
the resulting pressure ratio ($\sim 7.6^{+3.4}_{-2.0}$)
between the free-streaming ICM and
cluster gas at the stagnation point 
implies that NGC~4552 is 
moving supersonically (Mach $2.2^{+0.5}_{-0.3}$)
through the cluster gas with speed
$v \sim 1680^{+390}_{-220}$\kms at  
an angle $\xi \sim 35^{\circ} \pm 7^{\circ}$ towards us 
with respect to the plane of the sky.
}  

\item{The properties of the X-ray tail behind NGC~4552 are  
consistent with it being composed primarily of ram-pressure stripped 
galaxy gas. The tail is cool, with mean temperature 
$kT = 0.51^{+0.09}_{-0.06}$\,keV, and denser than the Virgo ICM, with
$n_{\rm et} \sim 5.4 \times 10^{-3}$\cmc  
for $\xi \sim 35^\circ$. 
Although the errors are large, mostly due  to uncertainties in the tail 
gas metallicity, the mean thermal pressure
of gas in the tail suggests that it is over-pressured with respect to the
ambient ICM. The subsequent adiabatic expansion of gas in the tail 
may partially explain the rapid fading of the tail with distance from 
the galaxy.
}

\end{enumerate}


\acknowledgements

This work is supported in part by NASA grant GO3-4176A  
and the Smithsonian Institution. 
This work has made use of the NASA/IPAC Extragalactic Database (NED)
which is operated by the Jet Propulsion Laboratory, California
Institute of Technology,  under contract with the National
Aeronautics and Space Administration. We wish to   
thank Maxim Markevitch for the use of his edge analysis codes.

\begin{small}

\end{small}

\begin{deluxetable}{ccccc}
\tablewidth{0pc}
\tablecaption{Spectral Analysis Regions Outside the Central 
$1.3$\,kpc of NGC~4552\label{tab:specregs}}
\tablehead{
\colhead{Region } &\colhead{Shape} &\colhead{Center}&
\colhead{Dimensions} &\colhead{Orientation}  \\
  &   & RA, DEC & arcsec & degrees }
\startdata
VN & rectangular &$12\,\,35\,\,37.7$, $12\,\,36\,\,16.5$ & $400$,$150$ &$319$ \\
VE & circular    &$12\,\,35\,\,55.5$, $12\,\,34\,\,33.7$ & $100$       & \\
NG & rectangular &$12\,\,35\,\,39.8$, $12\,\,33\,\,53.6$ & $13$, $42$ &$0$  \\
SG & rectangular &$12\,\,35\,\,39.8$, $12\,\,32\,\,52.9$ & $19$, $65$ &$0$  \\
OA & annular     &$12\,\,35\,\,39.7$, $12\,\,33\,\,22.9$ & $22$, $37$ &  \\
SL & circular    &$12\,\,35\,\,38.5$, $12\,\,32\,\,48.2$ & $17$       & \\
HW & elliptical  &$12\,\,35\,\,37.0$, $12\,\,33\,\,48.9$ & $30$, $12$ &$324$ \\
HE & elliptical  &$12\,\,35\,\,42.6$, $12\,\,33\,\,52.9$ & $28$, $10$ &$27$ \\
T  & rectangular &$12\,\,35\,\,42.6$, $12\,\,31\,\,57.6$ & $69$, $62$ &$52$\\
\enddata
\tablecomments{  WCS coordinates for the centers of the 
regions are J2000. Dimensions specified are radii for 
circular regions, (inner, outer) radii for annular regions,
(length,width) for rectangular regions and semi-(major, minor) 
axis for elliptical regions.  Orientation angles are for a region's 
major axis measured counterclockwise from west (+x axis in
Figure \protect\ref{fig:specimage}).  Source regions are also 
shown in Figure \protect\ref{fig:specimage} 
except for regions NG and SG, that were 
combined into the larger region OA to improve statistics.
HW and HE were combined into a single ``horn''  
region (HE+HW) in the spectral analysis to improve statistics for 
the horns. 
 }
\end{deluxetable}

\begin{deluxetable}{cccc}
\tablewidth{0pc}
\tablecaption{Background Regions\label{tab:bgsigregs}}
\tablehead{
\colhead{Region } &\colhead{Shape} &\colhead{Center}&
\colhead{Dimensions} \\
  &   & RA, DEC & arcsec }
\startdata
HE\_BG & circular &$12\,\,35\,\,35.7$, $12\,\,35\,\,34.2$ & $75.2$ \\
HW\_BG & circular &$12\,\,35\,\,35.7$, $12\,\,35\,\,34.2$ & $75.2$ \\
SL\_BG& elliptical  &$12\,\,35\,\,39.8$, $12\,\,33\,\,53.6$ & $20.7$,
$16$ \\
T\_BG &circular  &$12\,\,35\,\,52.4$, $12\,\,33\,\,19.4$ & $82.9$ \\
\enddata
\tablecomments{  Background regions used to compute the uncertainty in
  the flux measurement for regions HE, HW, R, SL, and T are denoted
as HE\_BG, HW\_BG, 
SL\_BG, and T\_BG, respectively. 
WCS coordinates for the centers of the 
regions are J2000. Dimensions specified are radii for 
circular regions, (inner, outer) radii for the annular region,
and semi-(major, minor)  
axis for the elliptical region.  The major axis of the elliptical region is
oriented at $0^\circ$ measured counterclockwise from west (+x axis in
Figure \protect\ref{fig:specimage}).  Source regions are listed in 
Table \protect\ref{tab:specregs}  
and are shown in Figure \protect\ref{fig:specimage}.  
 }
\end{deluxetable}

\begin{deluxetable}{ccccccc}
\tablewidth{0pc}
\tablecaption{Computation of the 
Precision of the Flux Measurement\label{tab:fluxsigdat}}
\tablehead{ 
\colhead{Region } &\colhead{Region} &\colhead{Region }&
\colhead{Bkgd } &\colhead{Bkgd} & \colhead{Bkgd} & \colhead{Source} \\
 ID & Cnts   & Area  & ID & Cnts & Area & Cnts }
\startdata
HE & $203$ & $864.4$ & HE\_BG & $931$ & $17716$  & $157 \pm 14$ \\
HW & $205$ & $1060$ & HW\_BG & $931$ & $17716$   & $149  \pm 14$ \\
SL & $744$ & $923.2$ & SL\_BG & $624$ & $985.7$  & $160 \pm 36$ \\
T  & $705$ & $4246$ & T\_BG & $1059$ & $21509$   & $495 \pm 27$ \\ 
\enddata
\tablecomments{ The columns represent (1) region
  identifier, (2) total counts in the $0.3-2$ keV energy band, (3)
  area of the region in arcsec$^2$, (4) background region identifier, 
(5) counts in the background region in the $0.3-2$\,keV energy band, 
(6) background region area in arcsec$^2$, and (7) the net source
counts and the uncertainty. The above counts and areas 
exclude contributions from point sources identified in each region. 
These are $53$ counts from
$2$ point sources ($12.58$\,arcsec$^2$) for region HE, $44$ counts from
$1$ point source ($6.29$\,arcsec$^2$) for region HW, $154$ counts from
$6$ point sources ($37.74$\,arcsec$^2$) for background region HE\_BG 
(HW\_BG), $32$ counts from $1$ point source
($6.29$\,arcsec$^2$) for region SL, $277$ counts from $8$ point sources
($50.32$\,arcsec$^2$) for region SL\_BG, $124$ counts from $1$ point source
($6.29$\,arcsec$^2$) for region T, and  $55$ counts from $8$ point sources
($84.45$\,arcsec$^2$) for region T\_BG.
 }
\end{deluxetable}

\begin{deluxetable}{ccccc}
\tablewidth{0pc}
\tablecaption{Best APEC Model Spectral Fits for Regions Outside 
the Central $1.3$\,kpc of NGC~4552\label{tab:allspectra}}
\tablehead{
\colhead{Region} &
\colhead{Source} &\colhead{$kT$} &\colhead{$A$}  
  &\colhead{$\chi^2/{\rm dof}$} \\
  & counts & keV & $\Zs$ & }
\startdata
VN  & $2113$ & $2.2^{+0.7}_{-0.4}$ &$0.1^{+0.2}_{-0.1}$ &$109/103$ \\
NG &$458 $  & $0.46^{+0.05}_{-0.04}$ & $0.2^{+0.2}_{-0.1}$   &$14.1/16$  \\
SG &$1070$  & $0.41^{+0.02}_{-0.03}$ & $0.5^{+0.9}_{-0.2}$   &$31/35$ \\
OA &$1990$  & $0.43^{+0.03}_{-0.02}$ & $0.4^{+0.2}_{-0.1}$ & $52/52$ \\
SL &$734$   & $0.40^{+0.03}_{-0.02}$& $ > 0.5$ & $39/26$ \\
SL$^{\dagger}$ & $734$ & $0.38 \pm 0.03$& $0.8 \pm 0.2$ &
$28/26$ \\
HE+HW &$346$ & $0.46^{+0.08}_{-0.06}$ & $0.2 \pm 0.1$&$25.7/15$\\
HE &$177$ & $0.39^{+0.08}_{-0.06}$ & $0.2^f$ & $4.4/8$ \\
HW &$155$ & $0.54 \pm 0.08$ & $0.2^f$ & $4/4$ \\
T &$517$  & $0.51^{+0.09}_{-0.06}$ &$0.4^{+1.1}_{-0.2}$ &$14.1/10$\\
T & $504$ & $0.51^{+0.09}_{-0.06}$ &$0.4^f$ & $14.1/11$ \\ 
\enddata
\tablecomments{ Column $2$ lists the background
subtracted source counts in the $0.3-7$\,keV energy band 
for region VN and in the $0.3-2$\,keV band for 
regions NG, SG, OA, SL, HE,HW, T.
Absorption is fixed at the Galactic value 
($n_{\rm H}=2.59 \times 10^{20}$\cms) in each 
region. Errors correspond to  $90\%$ confidence limits. 
Superscript $f$ denotes a fixed parameter. 
$\dagger$ VAPEC model with all abundances other than Fe
fixed at $0.4\Zs$  and $A_{\rm Fe}$ shown.
 }
\end{deluxetable}

\begin{deluxetable}{cccccc}
\tablewidth{0pc}
\tablecaption{Metallicity Dependence of Gas Outside the Rings\label{tab:metalscool}}
\tablehead{
\colhead{Region} &\colhead{$A$}& \colhead{$kT$}  
 & \colhead{$K$}&  \colhead{$\Lambda$}&\colhead{$\chi^2/{\rm dof}$} \\
  & $\Zs$& keV &$10^{-5}$\,cgs &$10^{-23}$erg\,cm$^{3}$s$^{-1}$ & }
\startdata
VN &      &                          &       &        &         \\ 
   & $0.1$  & $2.2^{+0.5}_{-0.4}$    &$20.3$ & $0.52$ &$109/104$ \\
   & $0.2$  & $2.3^{+0.6}_{-0.3}$    &$19.0$ & $0.55$ &$109/104$ \\
   & $0.3$  & $2.6 \pm 0.05$         &$17.8$ & $0.58$ &$111/104$ \\
OA &        &                        &       &        &         \\
   & $0.2$  & $0.44^{+0.03}_{-0.01}$ &$20.7$ & $0.5$  & $68.5/53$\\
   & $0.4$  & $0.43 \pm 0.02$        &$12.4$ & $0.9$  & $52.1/53$ \\
   & $0.5$  & $0.42^{+0.03}_{-0.02}$ &$10.4$ & $1.0$  &$53.5/53$ \\
   & $0.6$  & $0.42 \pm 0.02$        &$8.9$  & $1.2$  & $55/53$ \\
   & $0.8$  & $0.41^{+0.02}_{-0.01}$ &$6.9$  & $1.6$  & $59.5/53$ \\
   & $1.0$  & $0.41\pm 0.02$         &$5.6$  & $1.9$  & $62.8/53$ \\
T  &        &                        &       &        &         \\
   & $0.2$  & $0.54 \pm 0.07$        &$4.8$  & $0.6$  & $18.1/11$  \\
   & $0.4$  & $0.51^{+0.09}_{-0.06}$ &$2.9$  & $1.0$  & $14.1/11$ \\
   & $0.5$  & $0.5^{+0.1}_{-0.06}$   &$2.5$  & $1.2$  & $14.4/11$ \\
   & $0.6$  & $0.50^{+0.1}_{-0.06}$  &$2.1$  & $1.4$  & $14.5/11$  \\
   & $0.8$  & $0.49^{+0.1}_{-0.05}$  &$1.6$  & $1.8$  & $15.2/11$  \\
   & $1.0$  & $0.49^{+0.1}_{-0.06}$  &$1.3$  & $2.2$  & $15.7/11$  \\
HE+HW &     &                        &       &        &           \\
   & $0.1$  & $0.47^{+0.08}_{-0.07}$ & $5.3$ & $0.3$  & $27.1/16$ \\
   & $0.2$  & $0.46 \pm 0.07$        & $3.5$ & $0.5$  & $26.2/16$ \\
   & $0.3$  & $0.45 \pm 0.07$        & $2.5$ & $0.7$  & $28.2/16$ \\
   & $0.4$  & $0.45 \pm 0.07$        & $2.0$ & $0.9$  & $30.0/16$ \\
HE &     &                        &       &        &           \\
   & $0.1$  & $0.41^{+0.09}_{-0.07}$ & $3.1$ & $0.3$  & $6.3/8$ \\
   & $0.2$  & $0.39^{+0.08}_{-0.06}$ & $2.1$ & $0.5$  & $4.4/8$ \\
   & $0.3$  & $0.39 \pm 0.06$        & $1.5$ & $0.6$  & $4.8/8$ \\
   & $0.4$  & $0.37^{+0.08}_{-0.06}$ & $1.2$ & $0.8$  & $5.3/8$ \\
   &        &                        &       &        &         \\
HW &     &                        &       &        &           \\
   & $0.1$  & $0.54^{+0.09}_{-0.08}$ & $2.3$ & $0.4$  & $8.3/4$ \\  
   & $0.2$  & $0.54 \pm 0.08$        & $1.5$ & $0.6$  & $4/4$ \\
   & $0.3$  & $0.54^{+0.08}_{-0.09}$ & $1.1$ & $0.8$  & $2.8/4$ \\
   & $0.4$  & $0.54^{+0.08}_{-0.09}$ & $0.87$& $1.0$  & $2.3/4$ \\
\enddata
\tablecomments{ Dependence of the model temperature $kT$, APEC model
normalization $K$, and  
$0.5-2$\,keV X-ray emissivity $\Lambda$ 
on metal abundance $A$ for the nearby Virgo ICM, for gas in NGC~4552 
outside the rings ($r > 1.3$\,kpc) and gas in the horns and tail. 
The regions are defined in Table \protect\ref{tab:specregs} and best-fit 
model results listed in Table \protect\ref{tab:allspectra}. 
The abundance (column $2$) 
and hydrogen absorption ($n_{\rm H}=2.59 \times 10^{20}$\cms)  
are fixed, while the temperature and APEC model 
normalization are allowed to vary. Errors are $90\%$ confidence limits.
``cgs'' denotes cgs units for the APEC model normalization as given by XSPEC. 
 }
\end{deluxetable}

\begin{deluxetable}{cccccccc}
\tablewidth{0pc}
\tablecaption{Derived Velocities and Edge Properties for Different
Galaxy Abundances\label{tab:properties}}
\tablehead{
\colhead{$A$} & \colhead{$n_{e2}/n_{ec}$}&\colhead{$T_{2}/T_{c}$} & 
\colhead{$p_2/p_c$} & \colhead{$M$} & \colhead{$v$} 
 & \colhead{$v_{\rm t}$} & \colhead{$\xi$} \\
$\Zs$ & & & & & \kms & \kms & degrees }
\startdata
$0.2$ & $55$ & $0.2$ & $11.0$ &$2.7$ & $2070$ & $1830$ & $28$ \\
$0.4$ & $42$ & $0.2$ & $8.4$ & $2.3$ & $1760$ & $1470$ & $33$ \\
$0.5$ & $38$ & $0.2$ & $7.6$ & $2.2$ & $1680$ & $1380$ & $35$ \\ 
$0.6$ & $35$ & $0.2$ & $7.0$ & $2.1$ & $1610$ & $1290$ & $37$ \\
$0.8$ & $31$ & $0.2$ & $6.2$ & $2.0$ & $1530$ & $1190$ & $39$ \\
$1.0$ & $28$ & $0.2$ & $5.6$ & $1.9$ & $1460$ & $1090$ & $42$ \\
\enddata
\tablecomments{$n_e$, $T$, and $p$ denote the electron density, temperature
and pressure in each region specified by subscripts $2$ and $c$ 
for galaxy gas just inside the leading edge of the cold front 
and cluster gas in the free 
stream region, respectively (See Vikhlinin \etal 2001). $M$ is the Mach
number, $v$ is the total velocity, $v_{\rm t}$ is the
component of velocity in the plane of the sky, and $\xi$ is the 
inclination angle of motion towards the observer 
with respect to the plane of the
sky, respectively, for NGC~4552 relative to the Virgo ICM, 
 assuming a relative radial velocity between
NGC~4552 and the Virgo Cluster (M87) of $v_r = -967 \pm 11$\kms (NED;
Smith \etal 2000).
 }
\end{deluxetable}
\vfill
\eject
\end{document}